\documentclass[12pt, journal, onecolumn]{IEEEtran}
\usepackage{graphics}
\usepackage{graphicx}
\usepackage{amsfonts}
\usepackage{multirow}
\usepackage{epstopdf}
\usepackage{color}
\usepackage{float}
\usepackage{caption}
\usepackage{subcaption}
\usepackage{bm}
\usepackage{algorithm}
\usepackage{algorithmic}
\usepackage{amsmath}
\usepackage{pifont}
\usepackage[utf8]{inputenc}
\usepackage{mathtools}
\usepackage{amsthm}

\newcounter{myalgctr}

\newtheorem{defn}{Definition}

\newenvironment{assumption}{
   \refstepcounter{myalgctr}
   \textsc{{\bf A\themyalgctr}} 
   }

\DeclarePairedDelimiter\floor{\lfloor}{\rfloor}

\newcommand{\Tscr}{\mathcal T}
\newcommand{\Mscr}{\mathcal M}
\newcommand{\Vscr}{\mathcal V}
\newcommand{\Fscr}{\mathcal F}

\newcommand{\dsum}{\displaystyle\sum\limits}

\newcommand{\bbR}{\mathbb R}

\newcommand{\bbC}{\mathbb C}

\newcommand{\prox}{\textrm{prox}}

\newcommand{\diag}[1]{\textrm{diag}\left\{#1\right\}}

\DeclareMathOperator*{\argmin}{\arg\!\min}

\newcommand{\matst}[1]{{\fontfamily{qcr}\selectfont #1}}

\begin{document}
%
\title{SPARCOM: Sparsity Based Super-Resolution Correlation Microscopy}
%
%

\author{Oren Solomon,~\IEEEmembership{Student Member,~IEEE,}
        Yonina C. Eldar,~\IEEEmembership{Fellow,~IEEE,}
        Maor Mutzafi 
        and Mordechai Segev
\thanks{This project has received funding from the European Union's Horizon 2020 research and innovation program under grant agreement No. 646804-ERC-COG-BNYQ and from the Ollendorf Foundation.}
\thanks{O. Solomon (e-mail:  orensol@tx.technion.ac.il) and Y. C. Eldar (e-mail: yonina@ee.technion.ac.il) are with the Department of Electrical Engineering, Technion—Israel Institute of Technology, Haifa 32000.}
\thanks{M. Mutzafi (e-mail:  maormutz@tx.technion.ac.il) and M. Segev (e-mail: msegev@tx.technion.ac.il) are with the Department of Physics and Solid State Institute, Technion—Israel Institute of Technology, Haifa 32000.}}

%

\maketitle

\begin{abstract}
In traditional optical imaging systems, the spatial resolution  is limited by the physics of diffraction, which acts as a low-pass filter.
The information on sub-wavelength features is carried by evanescent waves, never reaching the camera, thereby posing a hard limit on resolution: the so-called diffraction limit. Modern microscopic methods enable super-resolution, by employing florescence techniques.
State-of-the-art localization based fluorescence sub-wavelength imaging techniques such as PALM and STORM achieve sub-diffraction spatial resolution of several tens of nano-meters. However, they require tens of thousands of exposures, which limits their temporal resolution. 
We have recently proposed SPARCOM ({\it sparsity based super-resolution correlation microscopy}), which exploits the sparse nature of the fluorophores distribution, alongside a statistical prior of uncorrelated emissions, and showed that SPARCOM achieves spatial resolution comparable to PALM/STORM, while capturing the data hundreds of times faster. Here, we provide 
a detailed mathematical formulation of SPARCOM, which in turn leads to an efficient numerical implementation, suitable for large-scale problems. We further extend our method to a general framework for sparsity based super-resolution imaging, in which sparsity can be assumed in other domains such as wavelet or discrete-cosine, leading to improved reconstructions in a variety of physical settings. 
\end{abstract}


%
\IEEEpeerreviewmaketitle

\section{Introduction}

Spatial resolution in diffractive optical imaging is limited by one half of the optical wavelength, known as Abbe's diffraction limit \cite{born2000principles, goodman2005intro}. Modern microscopic methods enable super-resolution, even though information on sub-wavelength features is absent in the measurements. One of the leading sub-wavelength imaging modalities is based on fluorescence (PALM \cite{Betzig2006b} and STORM \cite{rust2006sub}). Its basic principle consists of attaching florescent molecules (point emitters) to the features within the sample, exciting the fluorescence with short-wavelength illumination, and then imaging the fluorescent light. 
PALM and STORM rely on acquiring a sequence of diffraction-limited images, such that in each frame only a sparse set of emitters (fluorophores) are active. The position of each fluorophore 
is found through a super-localization procedure \cite{small2014fluorophore}. Subsequent accumulation of single-molecule localizations results in
 a grainy high-resolution image, which is then smoothed to form the final super-resolved image. 
 The final image has a spatial resolution of tens of nanometers. 

A major disadvantage of these florescence techniques is that they require tens of thousands of exposures. This is because in every frame, the diffraction-limited image of each emitter must be well separated from its neighbors, to enable the identification of its exact position. This inevitably leads to a long acquisition cycle, typically on the order of several minutes \cite{rust2006sub}. Consequently, fast dynamics cannot be captured by PALM/STORM. 

To reduce acquisition time, an alternative technique named SOFI (super-resolution optical fluctuation imaging) was proposed \cite{Dertinger2009}, which uses high fluorophore density, to reduce integration time. In SOFI, the emitters usually overlap in each frame, so that super-localization cannot be performed. However, since the emitted photons from each emitter, which are uncorrelated between different emitters, are captured over a period of several frames by the camera. Consecutive frames contain information in the pixel-wise temporal correlation between them. The measurements are therefore processed such that correlative information is used, enabling the recovery of features that are smaller than the diffraction limit by a factor of $\sqrt{2}$. By calculating higher order statistics (HOS) in the form of cumulants \cite{mendel1991tutorial} of the time-trace of each pixel, a theoretical resolution increase equal to the square root of the order of the statistics can in principle be achieved. Using the cross-correlation between pixels over time, 
it is possible to increase the resolution gain further, to an overall factor that scales linearly with the order of the statistical calculation \cite{Dertinger2010}.

SOFI enables processing of images with high fluorophore density, thus reducing the number of required frames for image recovery and achieving increased temporal resolution over localization based techniques. However, at least thus far, the spatial resolution offered by SOFI does not reach the level of super-resolution obtained through STORM and PALM, even when using HOS. The use of HOS can in principle increase the spatial resolution, but higher (than the order of two) statistical calculations require an increasingly large number of frames for their estimation, degrading temporal resolution. 
Moreover, SOFI suffers from a phenomenon known as {\it dynamic range expansion}, in which weak emitters are masked in the presence of strong ones. The effect is worsened as the statistical order increases, which in practice limits the applicability of SOFI to second order statistics and a moderate improvement in spatial resolution. 

Recently, we proposed a method for super-resolution imaging with short integration time called {\it sparsity based super-resolution correlation microscopy} (SPARCOM) \cite{Solomon2018sparsity}.  
In \cite{Solomon2018sparsity} we have shown that our method achieves spatial resolution similar to PALM/STORM, from only tens/hundreds of frames, by performing sparse recovery \cite{eldar2015sampling} on correlation information, leading to an improvement of the temporal resolution by two orders of magnitude. Mathematically, SPARCOM recovers the support of the emitters, by recovering their variance values. Sparse recovery from correlation information was previously proposed to improve sparse recovery from a small number of measurements \cite{PP2015, eldar2015sampling, Cohen2014}. When the non-zero entries of the sparse signal are uncorrelated, support size recovery can be theoretically increased up to $O(M^2)$, where $M$ is the length of a single measurement vector. In SPARCOM we use similar concepts to enhance resolution and improve the signal to noise ratio (SNR) in optical imaging. By performing sparse recovery on correlation information, SPARCOM enjoys the same features of SOFI (processing of high fluorophore density frames over short movie ensembles and the use of correlative information), while offering the possibility of achieving single-molecule resolution comparable to that of PALM/STORM. 
Moreover, by relying on correlation information only, SPARCOM overcomes the dynamic range problem of SOFI when HOS are used, and results in improved image reconstruction.

In this paper, we focus on three major contributions with respect to our recent work. The first is to provide a thorough and detailed formulation of SPARCOM, elaborating on its mathematical aspects. Second, we extend SPARCOM to the case when super-resolution is considered in additional domains such as the wavelet or discrete cosine transform domains. Third, we show how SPARCOM exploits structural information to achieve a computationally efficient implementation. This goal is achieved by considering the SPARCOM reconstruction model in the sampled Fourier space, which leads to fast image reconstruction, suitable for large-scale problems, without the need to store large matrices in memory. 

The rest of the paper is organized as follows: Section \ref{Sec:Problem} explains the problem and the key idea of SOFI. In Section \ref{Sec:FourierSOFI} we formulate our proposed solution. A detailed explanation of our algorithm, implementation and additional extensions to super-resolution in arbitrary bases are provided in Sections \ref{Sec:ProxGradSection} and \ref{Sec:Eff_imp}. Simulation results are presented in Section \ref{Sec:Sim}.

Throughout the paper, $x$ represents a scalar, ${\bf x}$ represents a vector, ${\bf X}$ a matrix and ${\bf I}_{N\times N}$ is the $N\times N$ identity matrix. The notation $||\cdot||_p$ represents the standard $p$-norm and $||\cdot||_F$ is the Frobenius norm. Subscript $x_l$ denotes the $l$th element of ${\bf x}$ and ${\bf x}_l$ is the $l$th column of ${\bf X}$. Superscript ${\bf x}^{(p)}$ represents ${\bf x}$ at iteration $p$, ${\bf T}^*$ denotes the adjoint of ${\bf T}$, and ${\bf \bar{A}}$ is the complex conjugate of ${\bf A}$.

\section{Problem formulation and SOFI}\label{Sec:Problem}
Following \cite{Dertinger2009, Dertinger2010}, the acquired fluorescence signal in the object plane is modeled as a set of $L$ independently fluctuating point sources, with resulting spatial fluorescence source distribution
    $$
        J({\bf r},t)=\sum_{k=0}^{L-1}\delta({\bf r} - {\bf r}_k) s_k(t).
    $$
    Each source (or emitter) has its own time dependent brightness function $s_k(t)$, and is located at position ${\bf r}_k\in\bbR^2,\;k=0,\ldots,L-1$. The acquired signal in the image plane is the result of the convolution between $J({\bf r},t)$ and the impulse response of the microscope $u({\bf r})$ (also known as the {\it point spread function} (PSF)),
    \begin{equation}
    \label{Eq:LowPassSig}
        f({\bf r},t)=\sum_{k=0}^{L-1}u({\bf r} - {\bf r}_k) s_k(t).
    \end{equation}
We assume that the measurements are acquired over a period of $t\in[0,T]$.  
    Ideally, our goal is to recover the locations of the emitters, ${\bf r}_k$ and their variances 
    with high spatial resolution and short integration time. The final high-resolution image is constructed from the recovered variance value for each emitter. 
    
    To proceed, we assume the following:\\
\begin{assumption}\label{as:1}
        The locations ${\bf r}_k,\;k=0,\ldots,L-1$ do not depend on time. 
\end{assumption}
\hspace{-0.75cm}
\begin{assumption}\label{as:2}
        The brightness is uncorrelated in space, namely, $E\{\tilde{s}_i(t_1)\tilde{s}_j(t_2)\}=0$, for all $i\neq j$, and for all $ t_1,t_2$, where $\tilde{s}_k(t)=s_k(t)-E_k$ with $E_k=E\{s_k(t)\}$. 
\end{assumption}
\hspace{-0.75cm}
\begin{assumption}\label{as:3}
The brightness functions $s_k(t),\;k=0,\ldots,L-1$ are wide sense stationary so that 
$E\{\tilde{s}_k(t)\tilde{s}_k(t+\tau)\}=g_{k}(\tau)$ for some function $g_{k}(\tau)$.
\end{assumption}

    Using assumptions {\bf A2} and {\bf A3}, 
    the autocorrelation function at each point ${\bf r}$ can be computed as
    \begin{equation}\label{Eq:SOFI}
      G_f({\bf r}, \tau)=E\{\tilde{f}({\bf r}, t)\tilde{f}({\bf r}, t+\tau)\}=\dsum_{k=0}^{L-1}u^2({\bf r}-{\bf r}_k) g_{k}(\tau), 
    \end{equation}
    where $\tilde{f}({\bf r},t)=f({\bf r},t)-E\{f({\bf r},t)\}=\sum_{k=0}^{L-1} u({\bf r}-{\bf r}_k)\tilde{s}_k(t)$. Assumption {\bf A1} indicates that ${\bf r}_k$ are time-independent during the acquisition period.
    The final SOFI image is the value of $G_f({\bf r}, 0)$ at each point ${\bf r}$, where $g_{k}(0)$ represents the variance of emitter $s_k$. We see from (\ref{Eq:SOFI}) that the autocorrelation function depends on the PSF squared. If the PSF is assumed to be Gaussian, then this calculation reduces its width by a factor of $\sqrt{2}$. However, the final SOFI image retains the same low resolution grid as the captured movie. Similar statistical calculations can be performed for adjacent pixels in the movie leading to a simple interpolation grid with increased number of pixels in the high-resolution image, but at the cost of increased statistical order using cumulants \cite{mendel1991tutorial}. HOS reduce the PSF size further but at the expense of degraded SNR and dynamic range for a given number of frames \cite{Dertinger2010}. 

    In the next section we provide a rigorous and detailed description of our sparsity based method, first presented in \cite{Solomon2018sparsity}, for estimating ${\bf r}_k$ and $g_{k}(0)$ on a high resolution grid. We rely on correlation only, without resorting to HOS, thus maintaining a short acquisition time, similar to correlation-based SOFI. In contrast to SOFI, we exploit the sparse nature of the emitters' distribution and recover a high-resolution image on a much denser grid than the camera's grid. This leads to spatial super-resolution without the need to perform interpolation using HOS \cite{Dertinger2010}. 


\section{SPARCOM}\label{Sec:FourierSOFI}
    \subsection{High resolution representation}
    To increase resolution by exploiting sparsity, we start by introducing a Cartesian sampling grid with spacing $\Delta_L$, which we refer to as the {\it low-resolution grid}. The low-resolution signal (\ref{Eq:LowPassSig}) can be expressed over this grid as
    \begin{equation}
    \label{Eq:Sampledf}
    \begin{array}{ll}
        f[m\Delta_L, n\Delta_L, t]=
        \dsum_{k=0}^{L-1}u[m\Delta_L-m_k,n\Delta_L-n_k]s_k(t),\;m,n=[0,\ldots,M-1],
    \end{array}
    \end{equation}
    where ${\bf r}_k=[m_k,n_k]^T\in\bbR^2$.
    We discretize the possible locations of the emitters ${\bf r}_k$, over a discrete Cartesian grid $i,l=0,\ldots,N-1$, $L\ll N$ with resolution $\Delta_H$, such that $[m_k,n_k]=[i_k,l_k]\Delta_H$ for some integers $i_k,l_k\in[0,\ldots,N-1]$. We refer to this grid as the {\it high-resolution grid}. For simplicity we assume that $\Delta_L=P\Delta_H$ for some integer $P\geq 1$, and consequently, $N = PM$. As each pixel $[m_k,n_k]$ is now divided into $P$ times smaller pixels, the high-resolution grid allows us to detect emitters with a spatial error which is $P$ times smaller than on the camera grid. Typical values of camera pixels sizes can be around $100$nm, which is typically half the diffraction limit. Thus, recovering the emitters on a finer grid leads to a better depiction of sub-diffraction features.  

    The latter discretization implies that (\ref{Eq:Sampledf}) is sampled (spatially) over a grid of size $M\times M$, while the emitters reside on a grid of size $N\times N$, with the $il$th pixel having a fluctuation function $s_{il}(t)$ (only $L$ such pixels actually contain fluctuating emitters, according to (\ref{Eq:Sampledf})). If there is no emitter in the $il$'th pixel, then $s_{il}(t)=0$ for all $t$. We further assume that the PSF $u$ is known.

    Rewriting (\ref{Eq:Sampledf}) in Cartesian form with respect to the grid of emitters yields
    \begin{equation}
    \label{Eq:FCS1}
    \hspace{-0.1cm}
    \begin{array}{ll}
       f[m\Delta_L,n\Delta_L,t]=
       \dsum_{i=0}^{N-1}\dsum_{l=0}^{N-1}u[m\Delta_L-i\Delta_H,n\Delta_L-l\Delta_H]s_{il}(t),
       \end{array}
    \end{equation}
    and additionally it holds that
    $$
    \begin{array}{ll}
        m\Delta_L-i\Delta_H=(mP-i)\Delta_H. \\
    \end{array}
    $$
    Omitting the spacing $\Delta_H$, we can rewrite (\ref{Eq:FCS1}) as
    \begin{equation}
    \label{Eq:FCS2}
        f[mP,nP,t]=\sum_{i,l=0}^{N-1}u[mP-i,nP-l]s_{il}(t).
    \end{equation}

    \subsection{Fourier analysis}
    We next present (\ref{Eq:FCS2}) in the Fourier domain, which will lead to an efficient implementation of our method.


    Since $y[m,n,t]=f[mP,nP,t]$ is an $M\times M$ sequence, denote by $Y[k_m,k_n,t]$ its $M\times M$ two dimensional discrete Fourier transform (DFT). 
    Performing an $M\times M$ two dimensional DFT on $y[m,n,t]$ yields
    \small$$\hspace{-0.0cm}
    \begin{array}{llll}
        Y[k_m,k_n,t]=\dsum_{m,n=0}^{M-1}f[mP,nP,t]e^{-j{2\pi\over M}k_m m}e^{-j{2\pi\over M}k_n n}=
        \dsum_{i,l=0}^{N-1}s_{il}(t)\dsum_{\hat{m},\hat{n}=0,P,\ldots}^{MP-P}u[\hat{m}-i,\hat{n}-l]e^{-j{2\pi\over MP}k_m \hat{m}}e^{-j{2\pi\over MP}k_n \hat{n}},
    \end{array}
    $$\normalsize
    where we defined $\hat{m}=mP$ and $\hat{n}=nP$ and $k_m,k_n=0,\ldots,M-1$.
    Next, consider $\hat{m},\hat{n}=0,\ldots,N-1$ and define the $N\times N$ sequence,
    \begin{equation}
    \label{Eq:psf_tbl}
        \tilde{u}[\hat{m},\hat{n}]=
        \left\{\begin{array}{ll}
        u\left[\hat{m},\hat{n}\right],\;\;\hat{m},\hat{n}=0,P,\ldots,N-P,\\
         \\
        0,\;\;\;\;\quad\quad\textrm{else},
        \end{array}\right.
    \end{equation}
    where $u$ is the discretized PSF sampled over $M\times M$ points of the low-resolution grid. We can then equivalently write 
    \begin{equation}
    \label{Eq:Y1}
    \begin{array}{ll}
        Y[k_m,k_n,t]=
        \dsum_{i,l=0}^{N-1}s_{il}(t)\dsum_{\hat{m},\hat{n}=0}^{N-1}\tilde{u}[\hat{m}-i,\hat{n}-l]e^{-j{2\pi\over N}k_m \hat{m}}e^{-j{2\pi\over N}k_n \hat{n}}.
    \end{array}
    \end{equation}
    By defining $p=\hat{m}-i$ and $q=\hat{n}-l$, (\ref{Eq:Y1}) becomes
    \begin{equation}
    \label{Eq:FCS3}
        Y[k_m,k_n,t]=\tilde{U}[k_m,k_n]\dsum_{i,l=0}^{N-1}s_{il}(t)e^{-j{2\pi\over N}k_m i}e^{-j{2\pi\over N}k_n l},
    \end{equation}
    with
    \begin{equation}
    \label{Eq:psf_f}
        \tilde{U}[k_m,k_n]=\dsum_{p,q=0}^{N-1}\tilde{u}[p,q]e^{-j{2\pi\over N}k_m p}e^{-j{2\pi\over N}k_n q}.
    \end{equation}
    
    Note that $\tilde{U}[k_m,k_n]$ 
    is the $N\times N$ two-dimensional DFT of the $N\times N$ sequence $\tilde{u}$, evaluated at discrete frequencies $k_m,k_n=0,\ldots,M-1$. From (\ref{Eq:psf_tbl}) and (\ref{Eq:psf_f}), it holds that $\tilde{U}[e^{-j{2\pi\over N }k_m},e^{-j{2\pi\over N }k_n}]=U[e^{-j{2\pi\over M }k_m},e^{-j{2\pi\over M}k_n}]$ for $k_m,k_n=0,\ldots,M-1$ ($N=PM$), where $U$ is the $M\times M$ two-dimensional DFT of $u$ sampled on the low-resolution grid. 



    Denote the column-wise stacking of each frame $Y[k_m,k_n,t]$ as an $M^2$ long vector ${\bf y}(t)$. In a similar manner, ${\bf s}(t)$ is a length-$N^2$ vector stacking of $s_{il}(t)$ for all $il$.  
    We further define the $M^2\times M^2$ diagonal matrix ${\bf H} = \diag{U[0,0],\ldots,U[M-1,M-1]}$. Vectorizing (\ref{Eq:FCS3}) yields
    \begin{equation}
    \label{Eq:yMat}
        {\bf y}(t)={\bf H}({\bf F}_M\otimes {\bf F}_M){\bf s}(t)={\bf A}{\bf s}(t),\;\;{\bf A}\in\bbC^{M^2\times N^2},
    \end{equation}
    where ${\bf s}(t)$ is an $L$-sparse vector and ${\bf F}_M$ denotes a partial $M\times N$ DFT matrix whose $M$ rows are the corresponding $M$ low frequency rows from a full $N\times N$ discrete Fourier matrix. 

    Define the autocorrelation matrix of ${\bf y}(t)$ as
    \small\begin{equation}
    \label{Eq:Ry}
        {\bf R}_{y}(\tau)=E\left\{({\bf y}(t)-E\{{\bf y}(t)\})({\bf y}(t+\tau)-E\{{\bf y}(t+\tau)\})^H \right\}.
    \end{equation}\normalsize
    From (\ref{Eq:yMat}),
    \begin{equation}
    \label{Eq:CorrY}
        {\bf R}_{y}(\tau)={\bf A}{\bf R}_{s}(\tau){\bf A}^H.
    \end{equation}
    Under assumption {\bf A2}, 
    ${\bf R}_{s}(\tau)$, the autocorrelation matrix of ${\bf s}(t)$, is a diagonal matrix. Therefore, (\ref{Eq:CorrY}) may be written as
    \begin{equation}
    \label{Eq:Ry_exp}
        {\bf R}_{y}(\tau)=\dsum_{l=1}^{N^2}{\bf a}_l{\bf a}_l^Hr_{{s}_{l}}(\tau),
    \end{equation}
    with ${\bf a}_l$ being the $l$th column of ${\bf A}$, ${\bf r}_s(\tau)=\diag{{\bf R}_s(\tau)}$, and $r_{{s}_{l}}(\tau)$ the $l$th entry of ${\bf r}_s(\tau)$. By taking $\tau=0$ we estimate the variance of $s_{ij}(t),\;i,j=0,\ldots,N-1$ (as written in assumption {\bf A3}). It is also possible to take into account the fact that the autocorrelation matrix ${\bf R}_y(\tau)$ may be non-zero for $\tau\neq 0$; for simplicity we use $\tau=0$.
    The support of ${\bf r}_{s}(\tau)$ is equivalent to the support of ${\bf s}(t)$, which in turn indicates the locations of the emitters on a grid with spacing $\Delta_H$. Thus, our high resolution problem reduces to recovering the $L$ non-zero values of $r_{{s}_{l}}(0)$ in (\ref{Eq:Ry_exp}). 

    \subsection{Sparse recovery}
    SPARCOM is based on (\ref{Eq:Ry_exp}), taking into account that ${\bf x}={\bf r}_s(0)$ is a sparse vector. We therefore find ${\bf x}$ by using a sparse recovery methodology. In our implementation of SPARCOM we use the LASSO formulation \cite{tibshirani1996regression} to construct the following convex optimization problem
    \begin{equation}
    \label{Eq:SFformulation}
        \min_{{\bf x}\geq {\bf 0}}\lambda||{\bf x}||_1+{1\over 2}\left|\left|{\bf R}_y(0) - \dsum_{l=1}^{N^2}{\bf a}_l{\bf a}_l^Hx_l\right|\right|_F^2,\tag{F-LASSO}
    \end{equation}
    with a regularization parameter $\lambda\geq 0$ and $x_l$ denoting the $l$th entry in ${\bf x}$. 
    We note that it is possible to write a similar formulation to (\ref{Eq:SFformulation}) accounting for $\tau>0$ (without the non-negativity constraint). Other approaches to sparse recovery may similarly be used.

    We solve (\ref{Eq:SFformulation}) iteratively using the FISTA algorithm \cite{palomar2010convex, Beck2009, Wimalajeewa2013}, which at each iteration performs a gradient step and then a thresholding step. By performing the calculations in the DFT domain, we can calculate the gradient of the smooth part of (\ref{Eq:SFformulation}), that is the squared Frobenius norm,
    very efficiently. We discuss this efficient implementation in detail in Section \ref{Sec:Eff_imp}.

    To achieve even sparser solutions, we implement a reweighted version of (\ref{Eq:SFformulation}) \cite{Candes2008},
    \small
    \begin{equation}
    \label{Eq:SFformulationR}
        {\bf x}^{(p+1)}=
        \argmin_{{\bf x}^{(p)}\geq {\bf 0}} \lambda||{\bf W}^{(p)}{\bf x}^{(p)}||_1+{1\over 2}\left|\left|{\bf R}_y(0) - \dsum_{l=1}^{N^2}{\bf a}_l{\bf a}_l^Hx_l^{(p)}\right|\right|_F^2
    ,
    \end{equation}\normalsize
    \normalsize
    where ${\bf W}$ is a diagonal weighting matrix and $p$ denotes the number of the current reweighting iteration. Starting from $p=1$ and ${\bf W}={\bf I}$, where ${\bf I}$ is the identity matrix of appropriate size, the weights are updated after a predefined number of FISTA iterations according to the output of ${\bf x}$ as
    $$
        W_{i}^{(p+1)}={1\over |x_i^{(p)}|+\epsilon},\;i=1,\ldots,N^2,
    $$
    where $\epsilon$ is a small non-negative regularization parameter. After updating the weights, the FISTA algorithm is performed again. 

    In practice, for a discrete time-lag $\tau$ and total number of frames $T$, ${\bf R}_{y}(\tau)$ is estimated from the movie frames using the empirical correlation
    $$
        {\bf R}_{y}(\tau)={1\over T-\tau}\dsum_{t=1}^{T-\tau}({\bf y}(t)-\bar{\bf y})({\bf y}(t+\tau)-\bar{\bf y})^H,
    $$
    with
    \begin{equation}
      \label{Eq:EmpAv}
      \bar{\bf y}={1\over T}\dsum_{t=1}^{T}{\bf y}(t).
    \end{equation}

    In the following sections we elaborate on our proposed algorithms for solving \ref{Eq:SFformulation} and the reweighted scheme (\ref{Eq:SFformulationR}). In particular, we explain how they can be implemented efficiently and extended to a more general framework of super-resolution under assumptions of sparsity. Table \ref{Tab:Tab1} provides a summary of the different symbols and their roles, for convenience.

\begin{table}
    \centering
    \caption{List of symbols}
    \label{tab:Hg}
    \begin{tabular}{ll}
        Symbol          & Description                                    \\
        $\otimes$ & Kronecker product                                       \\
        $\odot$ & Hadamard (element-wise) product\\
        M  & Number of pixels in one dimension of the low-resolution grid \\
        N  & Number of pixels in one dimension of the high-resolution grid \\
        $P$ & Ratio between $N$ and $M$ \\
    $\Delta_L$ & Low-resolution grid sampling interval \\
        $\Delta_H$ & High-resolution grid sampling interval \\
        $T$ & Number of acquired frames \\
        $L$ & Number of emitters in the captured sequence \\
        $m_k, n_k$ & Possible positions of emitters on the high-resolution Cartesian grid \\
        $L_f$ & Upper bound on the Lipschitz constant\\
        $\Tscr_{\alpha}(\cdot)$ & Soft thresholding operator with parameter $\alpha$ defined in (\ref{Eq:Prox})\\ 
        $\lambda$ & Regularization parameter\\
        $\mu$ & Smoothing parameter for Algorithm \ref{Alg:Alg_fpg_SM}\\
        $u(\cdot)$ & $M\times M$ discretized PSF\\
        ${\bf y}(t)$ & Vectorized $M\times M$ input frame at time $t$, after FFT\\
        ${\bf s}(t)$ & Vectorized $N\times N$ emitters intensity frame at time $t$\\
        ${\bf F}_M$ & Partial $M\times N$ DFT matrix of the $M$ lowest frequencies\\
        ${\bf H}$ & Diagonal $M^2\times M^2$ matrix containing the (vectorized) DFT of the PSF\\
        ${\bf A}$ & ${\bf A}={\bf H}({\bf F}_M\otimes {\bf F}_M)$, known $M^2\times N^2$ sensing matrix, as defined in (\ref{Eq:yMat}) \\
        ${\bf a}_i$ & $i$th column of ${\bf A}$\\
        $\bar{\bf y}$ & Empirical average of the acquired low-resolution frames defined in (\ref{Eq:EmpAv})\\
        ${\bf R}_y(\tau)$ & Auto-covariance matrix of input movie's pixels for time-lag $\tau$\\
        ${\bf R}_s(\tau)$ & Auto-covariance matrix of the emitters for time-lag $\tau$\\
        ${\bf r_s}/{\bf x}$ & Diagonal of ${\bf R}_s(\tau)$\\
        ${\bf M}$ & ${\bf M}=|{\bf A}^H{\bf A}|^2$\\
        ${\bf v}$ & ${\bf v}=[{\bf a}_1^H{\bf R}_y(0){\bf a}_1,\ldots,{\bf a}_{N^2}^H{\bf R}_y(0){\bf a}_{N^2}]^T$\\
        $\nabla f(\cdot)$ & Gradient of $f$ given by (\ref{Eq:grad_f})\\
        $K_{\textrm{max}}$ & Maximum number of iterations\\
        $\Mscr(\cdot)$ & Vector to matrix transformation, defined in (\ref{Eq:V2M})\\
        $\Vscr(\cdot)$ & Matrix to vector transformation, defined in (\ref{Eq:M2V})\\
    \end{tabular}\label{Tab:Tab1}
\end{table}

\section{Proximal gradient descent algorithms}
\label{Sec:ProxGradSection}
\subsection{Variance recovery}
    Problem (\ref{Eq:SFformulation}) can be viewed as a minimization of a decomposition model
    $$
        \min_{{\bf x}\geq{\bf 0}}\lambda g({\bf x})+f({\bf x}),
    $$
    where $f$ is a smooth, convex function with a Lipschitz continuous gradient and $g$ is a possibly non-smooth but proper, closed and convex function. Following \cite{Beck2009} and \cite{Wimalajeewa2013} we adapt a {\it fast-proximal algorithm}, similar to FISTA, to minimize the objective of (\ref{Eq:SFformulation}), as summarized in Algorithm \ref{Alg:Alg_fpg_lasso}. Solving (\ref{Eq:SFformulation}) iteratively involves finding {\it Moreau's proximal} (prox) mapping \cite{moreau1965proximite, tan2014smoothing} of $\alpha g$ for some $\alpha\geq 0$, defined as
    \begin{equation}
    \label{Eq:proxdef}
        \prox_{\alpha g}({\bf x})=\argmin_{{\bf u}\in \bbR^n}\left\{\alpha g({\bf u})+{1\over 2}||{\bf u}-{\bf x}||_2^2 \right\}.
    \end{equation}
    For $g(x)=||{\bf x}||_1$, $\prox_{\alpha g}({\bf x})$ is given by the well known {\it soft-thresholding} operator,
    \begin{equation}
    \label{Eq:Prox}
        \prox_{\alpha ||\cdot||_1}({\bf x})=\Tscr_{\alpha}({\bf x})=\max\{|{\bf x}|-\alpha, 0\}\cdot\textrm{sign}({\bf x}),
    \end{equation}
    where the multiplication, max and sign operators are performed element-wise. In its simplest form, the proximal-gradient method calculates the prox operator on the gradient step of $f$ at each iteration.

    Denoting
    \begin{equation}
    \label{Eq:f}
        f({\bf x})={1\over 2}\left|\left|{\bf R}_y(0) - \dsum_{l=1}^{N^2}{\bf a}_l{\bf a}_l^Hx_l\right|\right|_F^2,
    \end{equation}
    and differentiating it with respect to ${\bf x}$ yields
    \begin{equation}
    \label{Eq:grad_f}
      \nabla f({\bf x})={\bf M}{\bf x}-{\bf v},
    \end{equation}
    where ${\bf v}=[{\bf a}_1^H{\bf R}_y(0){\bf a}_1,\ldots,{\bf a}_{N^2}^H{\bf R}_y(0){\bf a}_{N^2}]^T$, ${\bf M}=|{\bf A}^H{\bf A}|^2$ and we have used the fact that ${\bf x}$ is real since it represents the variance of light intensities. The operation $|\cdot|^2$ is performed element-wise. The (upper bound on the) Lipschitz constant $L_f$ of $f({\bf x})$ is readily given by $L_f=||{\bf M}||_2$, corresponding to the largest eigenvalue of ${\bf M}$, since by (\ref{Eq:grad_f})
    $$
        ||\nabla f({\bf x})-\nabla f({\bf y})||_2\leq||{\bf M}||_2||{\bf x}-{\bf y}||_2.
    $$

    Calculation of (\ref{Eq:grad_f}) is the most computationally expensive part of Algorithm\footnote{Code is available at http://webee.technion.ac.il/people/YoninaEldar/software.php} \ref{Alg:Alg_fpg_lasso}. Since ${\bf M}$ is of dimensions $N^2\times N^2$, it is usually impossible to store it in memory and apply it straightforwardly in multiplication operations. In Section \ref{Sec:Eff_imp} we present an efficient implementation that overcomes this issue, by exploiting the structure of ${\bf M}$. We also develop a closed form expression for $L_f$.
    \begin{algorithm}
        \caption{Fast Proximal Gradient Descent for SPARCOM}
        \label{Alg:Alg_fpg_lasso}
        \begin{algorithmic}
            \REQUIRE $L_f$, ${\bf R}_y(0)$, $\lambda>0$, $K_{\textrm{max}}$
            \STATE {\bf Initialize} ${\bf z}_1={\bf x}_0={\bf 0}$, $t_1=1$ and $k=1$
            \WHILE {$k\leq K_{\textrm{max}}$ or stopping criteria not fulfilled}
                \STATE {\bf 1:} $\nabla f({\bf z}_k)={\bf M}{\bf z}_k-{\bf v}$
                \STATE {\bf 2:} ${\bf x}_k=\Tscr_{{\lambda\over L_f}}({\bf z}_k-{1\over L_f}\nabla f({\bf z}_k))$
                \STATE {\bf 3:} Project to the non-negative orthant ${\bf x}_k({\bf x}_k < {\bf 0}) = {\bf 0}$
                \STATE {\bf 4:} $t_{k+1}=0.5(1+\sqrt{1+4t_k^2})$
                \STATE {\bf 5:} ${\bf z}_{k+1}={\bf x}_k+{t_k-1\over t_{k+1}}({\bf x}_k-{\bf x}_{k-1})$
                \STATE {\bf 6:} $k\leftarrow k+1$
            \ENDWHILE
            \RETURN ${\bf x}_{K_{\text{max}}}$
        \end{algorithmic}
    \end{algorithm}

    Implementing the re-weighted $l_1$ minimization of (\ref{Eq:SFformulationR}) involves calculation of the following element-wise soft-thresholding operator
    \begin{equation}
    \label{Eq:Prox_re}
      \Tscr_{{\lambda\over L_f}W_{i}}(x_i)=\max\left\{|x_i|-{\lambda\over L_f}W_{i}, 0\right\}\cdot\textrm{sign}(x_i),
    \end{equation}
    with $W_{i}$ being the current value of the $i$th entry of the diagonal of the weighting matrix ${\bf W}$. The re-weighting procedure is summarized in Algorithm \ref{Alg:Alg_fpg_re}.
    \begin{algorithm}
        \caption{Iterative re-weighted Fast Proximal Gradient for (\ref{Eq:SFformulation})}
        \label{Alg:Alg_fpg_re}
        \begin{algorithmic}
            \REQUIRE $L_f$, ${\bf R}_y(0)$, $\lambda>0$, $\epsilon>0$, $P_{\textrm{max}}$
            \STATE {\bf Initialize} Set iteration counter $l=1$ and ${\bf W}^{1}={\bf I}$
            \WHILE {$p\leq P_{max}$ or stopping criteria not fulfilled}
                \STATE {\bf 1:} Solve (\ref{Eq:SFformulation}) using Algorithm \ref{Alg:Alg_fpg_lasso} with (\ref{Eq:Prox_re})
                \STATE {\bf 2:} Update weights for $i=1,\ldots,N^2$
                $$
                \begin{array}{ll}
                    W_{i}^{(p+1)}=\diag{{1\over \left|{\bf x}_1^{(p)}\right|+\epsilon},\ldots,{1\over \left|x_{N^2}^{(p)}\right|+\epsilon}},\\
                \end{array}
                $$
                \STATE {\bf 3:} $p\leftarrow p+1$
            \ENDWHILE
            \RETURN ${\bf x}_{P_{\textrm{max}}}$
        \end{algorithmic}
    \end{algorithm}

\subsection{Regularized super-resolution}
    Recall that to achieve super-resolution we assumed that the recovered signal is sparse. Such an assumption arises in the context of fluorescence microscopy, in which the imaged object is labeled with fluorescing molecules such that the molecular distribution or the desired features themselves are spatially sparse. 
In many cases the sought after signal has additional structure which can be exploited alongside sparsity, especially since attaching fluorescing molecules to sub-cellular organelles serves as means to image these structures, which are of true interest. Thus, when considering sparsity based super-resolution reconstruction, we can consider a more general context of sparsity within the desired signal.

    \subsubsection{Total variation super-resolution imaging}
        We first modify (\ref{Eq:SFformulation}) to incorporate a total-variation  regularization term on ${\bf x}$ \cite{rudin1992nonlinear, chambolle2004algorithm}, that is, we assume that the reconstructed super-resolved correlation-image is piece-wise constant:
        \begin{equation}
        \label{Eq:SFTV}
            \min_{{\bf x}\geq {\bf 0}}\lambda\textrm{TV}({\bf x})+{1\over 2}\left|\left|{\bf R}_y(0) - \dsum_{l=1}^{N^2}{\bf a}_l{\bf a}_l^Hx_l\right|\right|_F^2.\tag{F-TV}
        \end{equation}
        We follow the definition of the discrete $\textrm{TV}({\bf x})$ regularization term as described in \cite{beck2009fast}, for both the isotropic and anisotropic cases. The proximity mapping $\prox_{\alpha \textrm{TV}}({\bf x})$ does not have a closed form solution in this case. Instead, the authors of \cite{beck2009fast} proposed to solve $\prox_{\alpha \textrm{TV}}({\bf x})$ iteratively. The minimizer of (\ref{Eq:proxdef}) is the solution to a {\it denoising} problem with the regularizer $\alpha g(\cdot)$ on the recovered signal. In particular, $\prox_{\alpha \textrm{TV}}({\bf x})$ is the denoising solution with total-variation regularization. Many total-variation denoising algorithms exist (e.g. \cite{rudin1992nonlinear, chambolle2004algorithm, osher2005iterative} and \cite{figueiredo2006total}), thus any one of them can be used to calculate the proximity mapping iteratively. In particular, we chose to follow the fast TV denoising method suggested in \cite{beck2009fast} and denoted as Algorithm GP. The algorithm accepts an observed image, a regularization parameter $\lambda$ which balances between the level of sparsity and compatibility to the observations and a maximal number of iterations $N_{max}$. The output is a TV denoised image. Thus, as summarized in Algorithm \ref{Alg:Alg_fpg_TV}, each iterative step is composed of a gradient step of $f$ and a subsequent application of Algorithm GP.

        Algorithm GP already incorporates a projection onto box constraints, which also includes as a special case the non-negativity constraints of (\ref{Eq:SFTV}). Hence we have omitted the projection step in Algorithm \ref{Alg:Alg_fpg_TV}. 


        \begin{algorithm}
        \caption{Fast Proximal Gradient Descent for (\ref{Eq:SFTV})}
        \label{Alg:Alg_fpg_TV}
        \begin{algorithmic}
            \REQUIRE $L_f$, ${\bf R}_y(0)$, $\lambda>0$, $K_{\textrm{max}}$, $N_{\textrm{max}}$
            \STATE {\bf Initialize} ${\bf z}_1={\bf x}_0={\bf 0}$, $t_1=1$ and $k=1$
            \WHILE {$k\leq K_{\textrm{max}}$ or stopping criteria not fulfilled}
                \STATE {\bf 1:} $\nabla f({\bf z}_k)={\bf M}{\bf z}_k-{\bf v}$
                \STATE {\bf 2:} ${\bf x}_k=\textrm{GP}({\bf z}_k-{1\over L_f}\nabla f({\bf z}_k), \lambda, N_{\textrm{max}})$
                \STATE {\bf 3:} $t_{k+1}=0.5(1+\sqrt{1+4t_k^2})$
                \STATE {\bf 4:} ${\bf z}_{k+1}={\bf x}_k+{t_k-1\over t_{k+1}}({\bf x}_k-{\bf x}_{k-1})$
                \STATE {\bf 5:} $k\leftarrow k+1$
            \ENDWHILE
            \RETURN ${\bf x}_{K_{\textrm{max}}}$
        \end{algorithmic}
        \end{algorithm}



    \subsubsection{Analysis type super-resolution imaging}
        In many scenarios, additional priors can be exploited alongside sparsity, to achieve sub-wavelength resolution. 
        Examples include wavelet transforms and the discrete cosine transforms (DCT). In general, the problem we wish to solve is
        $$
            \min_{{\bf x}}\lambda||{\bf T}^*{\bf x}||_1+{1\over 2}\left|\left|{\bf R}_y(0) - \dsum_{l=1}^{N^2}{\bf a}_l{\bf a}_l^Hx_l\right|\right|_F^2,
        $$
        Where ${\bf T}\in\bbC^{M\times N}$ is some known transformation. 
        The prox mapping of the regularization term $||{\bf T}^*{\bf x}||_1$ does not admit a closed form solution. The authors of \cite{tan2014smoothing} suggested to approximate the generally non differentiable function $f({\bf x})+g({\bf T}^*{\bf x})$ with a surrogate differentiable function, thus alleviating the need to calculate the prox mapping of the non-differentiable term $g({\bf T}^*{\bf x})$. The smooth surrogate function used is the {\it Moreau envelope of g} \cite{moreau1965proximite}, given by
        $$
            g_{\mu}({\bf x})=\min_{{\bf u}}\left\{g({\bf u})+{1\over 2\mu}||{\bf u}-{\bf x}||_2^2\right\}.
        $$
        
        We therefore propose a smooth counterpart to (\ref{Eq:SFformulation}),
        \begin{equation}
        \label{Eq:SF_Smoothed}
            \min_{{\bf x}\geq{\bf 0}}f({\bf x})+g_{\mu}({\bf T}^*{\bf x}),\tag{F-SM}
        \end{equation}
        with $f({\bf x})$ given by (\ref{Eq:f}) 
        and $g_{\mu}({\bf x})$ given by
        $$
            g_{\mu}({\bf T}^*{\bf x})=\min_{{\bf u}}\left\{\lambda||{\bf u}||_1+{1\over 2\mu}||{\bf u}-{\bf T}^*{\bf x}||_2^2\right\}.
        $$
        The gradient of (\ref{Eq:SF_Smoothed}) is now a combination of the gradients of $f({\bf x})$ and $g_{\mu}({\bf x})$, with
        \begin{equation}
        \label{Eq:gmu}
            \nabla g_{\mu}({\bf T}^*{\bf x})={1\over\mu}{\bf T}({\bf T}^*{\bf x}-\Tscr_{\lambda\mu}({\bf T}^*{\bf x})).
        \end{equation}
        Using (\ref{Eq:gmu}) we have modified the SFISTA algorithm in \cite{tan2014smoothing} to solve (\ref{Eq:SF_Smoothed}), as summarized in Algorithm \ref{Alg:Alg_fpg_SM}. Note that the Lipschitz constant of $f({\bf x})+g_{\mu}({\bf T}^*{\bf x})$ is given by $L_f+{||{\bf T}||_2^2\over \mu}$. 
        \begin{algorithm}
        \caption{Fast Proximal Gradient descent for (\ref{Eq:SF_Smoothed})}
        \label{Alg:Alg_fpg_SM}
        \begin{algorithmic}
            \REQUIRE $L_f$, $\mu$, ${\bf R}_y(0)$, $\lambda>0$, $K_{\textrm{max}}$
            \STATE {\bf Initialize} ${\bf z}_1={\bf x}_0={\bf 0}$, $t_1=1$ and $k=1$
            \WHILE {$k\leq K_{\textrm{max}}$ or stopping criteria not fulfilled}
                \STATE {\bf 1:} $\nabla f({\bf z}_k)={\bf M}{\bf z}_k-{\bf v}$
                \STATE {\bf 2:} $\nabla g_{\mu}({\bf T}^*{\bf x}_{k-1})={1\over\mu}{\bf T}({\bf T}^*{\bf x}_{k-1}-\Tscr_{\lambda\mu}({\bf T}^*{\bf x}_{k-1}))$
                \STATE {\bf 3:} ${\bf y}_k={\bf z}_k-{1\over L_f}(\nabla f({\bf z}_k)+\nabla g_{\mu}({\bf T}^*{\bf x}_{k-1}))$
                \STATE {\bf 4:} $t_{k+1}=0.5(1+\sqrt{1+4t_k^2})$
                \STATE {\bf 5:} ${\bf z}_{k+1}={\bf x}_k+{t_k-1\over t_{k+1}}({\bf x}_k-{\bf x}_{k-1})+{t_k\over t_{k+1}}({\bf y}_k-{\bf x}_k)$
                \STATE {\bf 6:} $k\leftarrow k+1$
            \ENDWHILE
            \RETURN ${\bf x}_{K_{\textrm{max}}}$
        \end{algorithmic}
        \end{algorithm}



\section{Efficient Implementation}
    \label{Sec:Eff_imp}
        Solving (\ref{Eq:SFformulation}), (\ref{Eq:SFTV}) and (\ref{Eq:SF_Smoothed}) in practice can be very demanding in terms of numerical computations, due to the large dimensions of the reconstructed super-resolved image. Consider for example an input movie with frames of size $64\times 64$ pixels and a reconstructed super-resolved image of size $512\times 512$ pixels (an eight-fold increase in the density of the high-resolution grid compared to the low-resolution captured movie). Calculating ${\bf R}_y(0)$ yields a covariance data matrix of size $64^2\times 64^2$ and ${\bf R}_s(0)$ is of size $512^2\times 512^2$ pixels (though in practice its a diagonal matrix with a diagonal of length $512^2$ pixels). The exponential growth in the problem dimensions on the one hand and the diagonal structure of the covariance matrix of the super-resolved image on the other, prompts the search for an efficient implementation for Algorithms \ref{Alg:Alg_fpg_lasso}-\ref{Alg:Alg_fpg_SM}. We now show that by considering the signal model in the spatial frequency domain as in (\ref{Eq:FCS3}), an efficient implementation based on FFT and IFFT operations is possible.

    \subsection{Frequency domain structure}
        Recall that
        $$
            \nabla f({\bf x})={\bf M}{\bf x}-{\bf v},
        $$
        with ${\bf M}=|{\bf A}^H{\bf A}|^2$, ${\bf A}={\bf H}({\bf F}_M\otimes {\bf F}_M)$ and ${\bf v}=[{\bf a}_1^H{\bf R}_y(0){\bf a}_1,\ldots,{\bf a}_{N^2}^H{\bf R}_y(0){\bf a}_{N^2}]^T$. Reconstruction of a super-resolved image of size ${N\times N}$ dictates that ${\bf M}$ will be of size $N^2\times N^2$. In most cases, it is impossible to store ${\bf M}$ in memory or perform matrix-vector multiplications. Instead, we exploit the special structure of ${\bf M}$ to achieve efficient matrix-vector operations without explicitly storing it. 

\begin{figure}[ht!]
          \begin{subfigure}[b]{0.5\linewidth}
            \centering
            \includegraphics[trim={2.5cm 0cm 2.5cm 0cm},clip,width=0.8\linewidth]{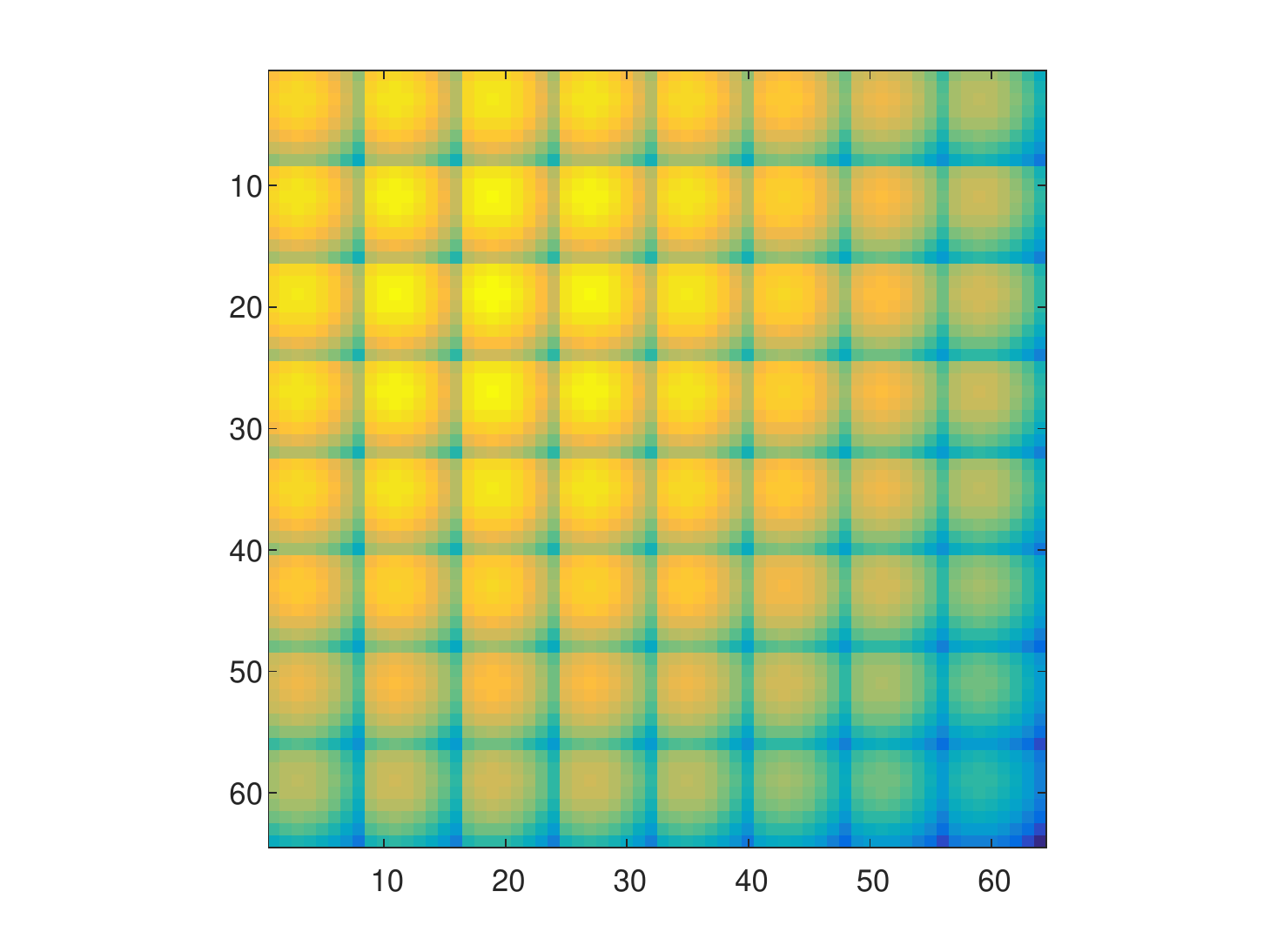}
            \caption{Image space domain.}\label{Fig:MtrU}
            \vspace{4ex}
          \end{subfigure}
          \begin{subfigure}[b]{0.5\linewidth}
            \centering
            \includegraphics[trim={2.5cm 0cm 2.5cm 0cm},clip,width=0.8\linewidth]{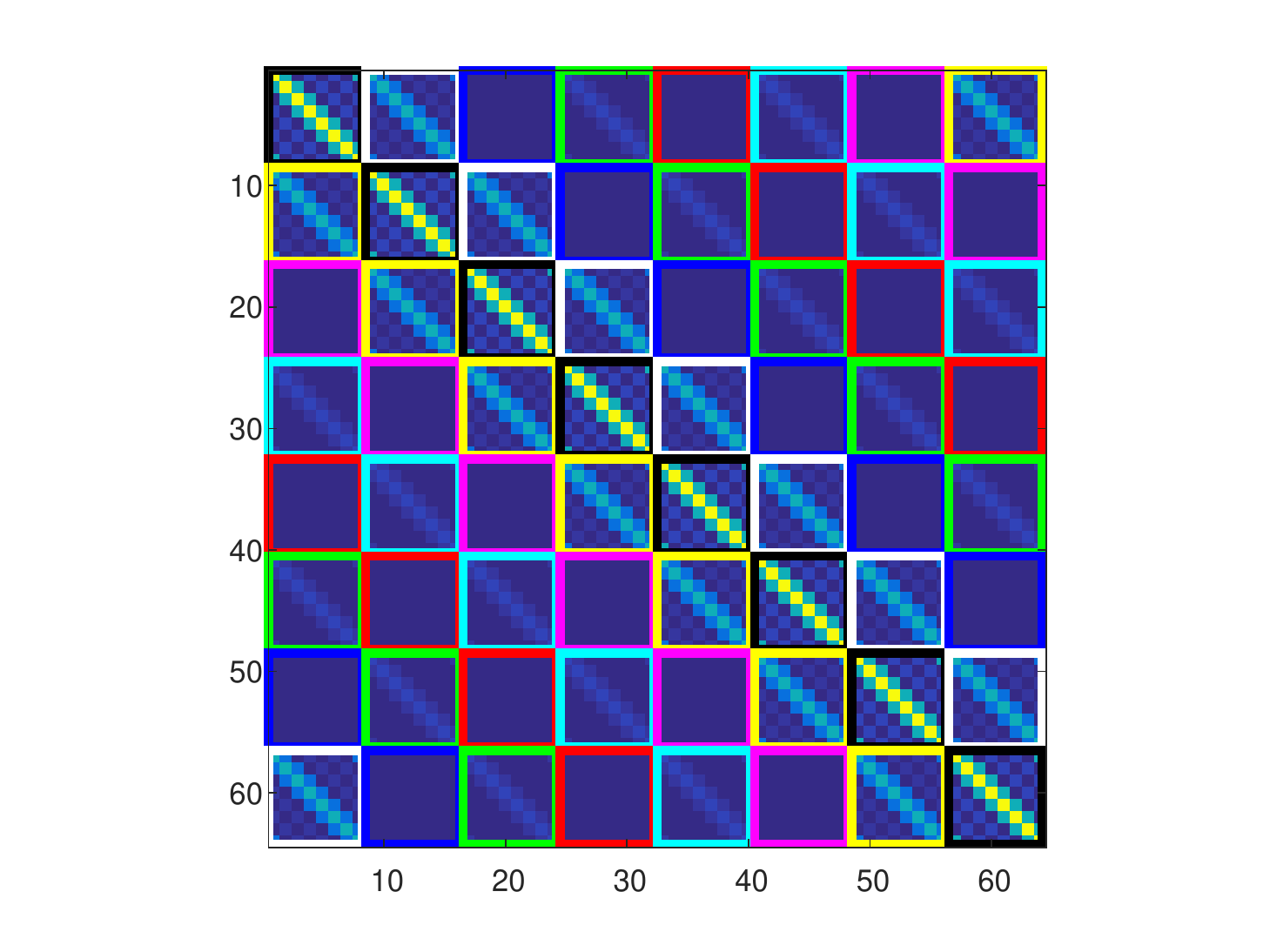}
            \caption{Discrete Fourier domain.}\label{Fig:MtrF}
            \vspace{4ex}
          \end{subfigure}\vspace{-0.5cm}
          \caption{Comparison of the structure of ${\bf M}$ for two possible formulations: Image space domain (a) and spatial discrete frequency domain (b).}\label{Fig:MtrComp}
        \end{figure}

        Figure \ref{Fig:MtrComp} illustrates the structure of ${\bf M}$ for two formulations: Fig. \ref{Fig:MtrComp}a shows the structure of ${\bf M}$ if we do not consider performing an FFT on (\ref{Eq:FCS2}). In this case, the $n$th column of ${\bf A}$ contains the vectorized $M\times M$ PSF centered at the high resolution pixel $ij,\;i,j\in[0,\ldots,N-1]$. Figure \ref{Fig:MtrComp}b illustrates the structure of ${\bf M}$ in the spatial-frequency domain, derived from (\ref{Eq:FCS3}). In both cases, ${\bf M}$ is of size $64\times 64$ pixels and is generated from a PSF of size $4\times 4$ pixels. Both figures represent a reconstruction of a $64\times 64$ super-resolved image.

        Figure \ref{Fig:MtrComp}b implies that the spatial-frequency formulation of ${\bf M}$ has a cyclic structure. This special structure will play a crucial role in our algorithm, as it  leads to efficient implementation of matrix vector multiplications. More specifically, in Appendix \ref{Seq:append} we show that ${\bf M}$ is {\it block circulant with circulant blocks} (BCCB) \cite{hansen2006deblurring}. Figure \ref{Fig:MtrComp}b can be divided into $8\times 8$ blocks (the different blocks are marked with rectangles of different colors to illustrate the block circular structure of the matrix), each block of size $8\times 8$ pixels. As can be seen, ${\bf M}$ is circulant with respect to the blocks and each block is also circulant. 

        Similar to circulant matrices which are diagonalizable by the DFT matrix, BCCB matrices are diagonalizable by the Kronecker product of two DFT matrices of appropriate dimensions. Such structure allows the implementation of a fast matrix-vector multiplication using FFT and inverse FFT operations without the need to store ${\bf M}$ in memory. In the following sections we describe the implementation of (\ref{Eq:grad_f}) in detail, by defining several operators which play a crucial role in its  calculation.

        \subsection{Efficient Implementation of ${\bf M}{\bf x}$}
        \label{Sec:EffImpM}
        We first define $\Mscr({\bf x})$, which takes ${\bf x}\in\bbC^{N^2}$ and transforms it into a matrix ${\bf X}\in\bbC^{N\times N}$, that is
        \begin{equation}
        \label{Eq:V2M}
            {\bf X}=\Mscr({\bf x}).
        \end{equation}
        This operation is performed using a column-wise division from top to bottom of ${\bf x}$. Upon dividing ${\bf x}$ into $N$ sub-vectors of length $N$ each, the $i$th column of ${\bf X}$ corresponds to the $i$th sub-vector of ${\bf x}$. Similarly, we denote the vectorization of ${\bf X}\in\bbC^{N\times N}$ which stacks the columns of ${\bf X}$ by 
        \begin{equation}
        \label{Eq:M2V}
            {\bf x}=\textrm{vec}({\bf X})=\Vscr({\bf X}).
        \end{equation}
        Here, ${\bf x}$ is a vector of length $N^2$, whose $i$th sub-vector of length $N$ corresponds to the $i$th column of ${\bf X}$.

        In Appendix \ref{Seq:append} we show that ${\bf M}$ is an $N^2\times N^2$ BCCB matrix with blocks of size $N\times N$. It is well known that such a matrix is diagonalizable by the Kronecker product of two discrete $N\times N$ Fourier matrices ${\bf F}_2={\bf F}\otimes {\bf F}$ \cite{hansen2006deblurring}, so that
        \begin{equation}
        \label{Eq:BCCB1}
            {\bf M}={\bf F}_2^H{\bm\Lambda}{\bf F}_2
        \end{equation}
        with ${\bm\Lambda}$ a diagonal matrix containing the eigenvalues of ${\bf M}$ on its diagonal. 
        To compute ${\bf M}{\bf x}$ we therefore need to calculate the eigenvalues of ${\bf M}$, and apply ${\bf F}_2$ and ${\bf F}_2^H$ on a given vector.

        Now,
        \begin{equation}
        \label{Eq:F2x}
            {\bf F}_2{\bf x}=({\bf F}\otimes {\bf F}){\bf x}=\Vscr({\bf F}\Mscr({\bf x}){\bf F}^T).
        \end{equation}
        The matrix ${\bf F}\Mscr({\bf x}){\bf F}^T$
        corresponds to applying the FFT on each column of $\Mscr({\bf x})$, and then again over the rows of the result. 
        In MATLAB, ${\bf F}_2{\bf x}$ is easily performed by reshaping ${\bf x}$ to $\Mscr({\bf x})$, applying the \matst{fft2} command on $\Mscr({\bf x})$ and vectorizing the result. Similarly, calculation of the 2D inverse FFT of an $N\times N$ matrix ${\bf X}_f$ is equivalent to ${1\over N^2}{\bf F}^H{\bf X}_f{\bf \bar{F}}$
        and is easily implemented in MATLAB with the \matst{ifft2} command. To compute the eigenvalues of ${\bf M}$ efficiently, we first need to be able to compute ${\bf A}{\bf x}$ and ${\bf A}^H{\bf x}$ for some ${\bf x}\in\bbC^{N^2}$.

        \subsubsection{Calculation of ${\bf A}{\bf x}$}
        Recall that
        $$
            {\bf A} = {\bf H}({\bf F}_M\otimes {\bf F}_M),
        $$
        where ${\bf F}_M\in\bbC^{M\times N}$ denotes a partial Fourier matrix, corresponding to the low-pass values of a full $N\times N$ Fourier matrix. 
        The operator ${\bf A}{\bf x}$ corresponds to taking ${\bf X}=\Mscr({\bf x})$, calculating ${\bf F}_M{\bf X}{\bf F}_M^T$, vectorizing the result and multiplying by ${\bf H}$. Denote,
        \begin{equation}
        \label{Eq:pfft2}
            \Fscr_{M_2}({\bf X})={\bf F}_M{\bf X}{\bf F}_M^T.
        \end{equation}
        The application of ${\bf F}_M$ on an $N\times N$ matrix ${\bf X}$ can be implemented by computing an FFT on each column of ${\bf X}$ and taking only the first $M$ rows of the result. Similarly, calculation of ${\bf X}{\bf F}_M^T=({\bf F}_M{\bf X}^T)^T$ is achieved by performing an FFT on each row of ${\bf X}$, taking the first $M$ rows of the result and performing the transpose operation. 

        Equation (\ref{Eq:pfft2}) implements a partial 2D-FFT operation on ${\bf X}$, where the full 2D-FFT operation is written as ${\bf F}{\bf X}{\bf F}^T$ with an $N\times N$ discrete Fourier matrix ${\bf F}$. The multiplication ${\bf A}{\bf x}$ can then be summarized as follows,
        \begin{equation}
        \label{Eq:AX}
            {\bf A}{\bf x} = {\bf H}\cdot \Vscr(\Fscr_{M_2}(\Mscr({\bf x}))).
        \end{equation}
        Since ${\bf H}$ is a diagonal matrix, the matrix-vector multiplication in (\ref{Eq:AX}) corresponds simply to multiplying the diagonal of ${\bf H}$ and the corresponding vector. If instead of a vector ${\bf x}$ we perform ${\bf A}{\bf Z}$ on some matrix ${\bf Z}\in\bbC^{N^2\times L}$, then the operation is performed on each column of ${\bf Z}$.


        \subsubsection{Calculation of ${\bf A}^H{\bf x}$}
        For ${\bf x}\in\bbC^{M^2}$,
        $$
            {\bf A}^H{\bf x}=({\bf F}_M^H\otimes {\bf F}_M^H){\bf H}^H{\bf x}=({\bf F}_M^H\otimes {\bf F}_M^H){\bf z}.
        $$
        Upon reformulating ${\bf z}$ as an $M\times M$ matrix ${\bf Z} = \Mscr({\bf z})$, we have
        $$
            \Mscr({\bf A}^H{\bf x})={\bf F}_M^H{\bf Z}{\bf \bar{F}}_M.
        $$
        Since ${\bf F}^H=N{\bf F}^{-1}$, ${\bf F}_M^H{\bf Z}$ corresponds to performing an inverse FFT on the zero-padded columns of ${\bf Z}$ and multiplying by $N$. We denote the result as ${\bf Y}$. Next, notice that ${\bf Y}{\bf \bar{F}}_M=({\bf F}_M^T{\bf Y}^H)^H$. Since the DFT matrix is a symmetric matrix, the second step involves computing an FFT on the zero-padded columns of ${\bf Y}^H$ and finally, taking the Hermitian operation.
        By denoting $\Fscr_{M_2}^H({\bf X})={\bf F}_M^H{\bf X}{\bf \bar{F}}_M$
         we can write
        \begin{equation}
        \label{Eq:AHx}
            {\bf A}^H{\bf x}= \Vscr(\Fscr_{M_2}^H(\Mscr({\bf H}^H{\bf x}))).
        \end{equation}
        If instead of a vector ${\bf x}$ we perform ${\bf A}^H({\bf Q})$ on some matrix ${\bf Q}\in\bbC^{M^2\times L}$, then the operation is performed on each column of ${\bf Q}$.


        \subsubsection{Calculation of the eigenvalues of ${\bf M}$}        
         To calculate the eigenvalues of ${\bf M}$, denoted by ${\bm \lambda}$, note that from (\ref{Eq:BCCB1})  
         ${1\over N}{\bf F}_2{\bf M}={\bm\Lambda} {\bf F}_2$, which implies that ${1\over N}{\bf F}_2{\bf m}_1={\bm\Lambda} {\bf f}_1$ with ${\bf m}_1$ and ${\bf f}_1$ being the first columns of ${\bf M}$ and ${\bf F}_2$, respectively. Since ${\bf f}_1$ is a vector of ones, we have
        $$
            {1\over N}{\bf F}_2{\bf m}_1={\bm \lambda},
        $$
        with ${\bm \lambda}=\diag{{\bm\Lambda}}$. To compute ${\bf m}_1$ we note that since ${\bf M}=|{\bf A}^H{\bf A}|^2$, ${\bf m}_1=|{\bf A}^H{\bf a}_1|^2$ where ${\bf a}_1$ 
        is the first column of ${\bf A}$. From the definition of ${\bf A}$, ${\bf a}_1={\bf h}$, where ${\bf h}=\diag{\bf H}$, and therefore ${\bf m}_1 = \left|({\bf F}_M^H\otimes{\bf F}_M^H)|{\bf h}|^2\right|^2$. In MATLAB, this can be implemented using \matst{fft} / \matst{ifft} operations, as noted by the first two steps of Algorithm \ref{Alg:Alg6}. By denoting the $M\times M$ DFT of the PSF as ${\bf U}$, it follows straightforwardly that $\Mscr({\bf m}_1)={\bf Z}=\left|\Fscr_{M_2}^H(|{\bf U}|^2)\right|^2$, where the operation $|\cdot|^2$ is performed element-wise. After the calculation of ${\bf m}_1$, finding ${\bm\lambda}$ is straightforward, since
        $$
            \Mscr({\bm\lambda})={\bf B}={\bf F}{\bf Z}{\bf F}^T,
        $$
        which can be computed using the 2D-FFT.

        We can summarize the application of ${\bf M}$ on ${\bf x}$ in Algorithm \ref{Alg:Alg6}, with ${\bf A}\odot {\bf B}$ representing the Hadamard element-wise product of two matrices ${\bf A}$ and ${\bf B}$, and the \matst{fft} / \matst{ifft} operations performed columnwise.

        \begin{algorithm}
            \caption{Calculation of ${\bf M}{\bf x}$}
            \label{Alg:Alg6}
            \begin{algorithmic}
                \REQUIRE The DFT of the PSF ${\bf U}$ and ${\bf x}$
                \STATE {\bf Eigenvalues calculation:}
                \STATE \hspace{0.3cm}{\bf 1:} Calculate ${\bf T}=N$\matst{ifft}$\{|{\bf U}|^2\}$ of length $N$
                \STATE \hspace{0.3cm}{\bf 2:} Calculate ${\bf E}=$\matst{fft}$\{{\bf T}^H\}$ of length $N$
                \STATE \hspace{0.3cm}{\bf 3:} Eigenvalues calculation ${\bf B}={\bf F}|{\bf E}^H|^2{\bf F}^T$ using \matst{fft2} 
                \STATE {\bf Application of ${\bf M}{\bf x}$:}
                \STATE \hspace{0.3cm}{\bf 4:} Calculate ${\bf Q}={\bf B}\odot ({\bf F}\Mscr({\bf x}){\bf F}^T)$
                \STATE \hspace{0.3cm}{\bf 5:} Calculate ${\bf Y}={1\over N^2}{\bf F}^H{\bf Q}{\bf \bar{F}}$ using \matst{ifft2} 
                \RETURN $\Vscr({\bf Y})={\bf M}{\bf x}$.
            \end{algorithmic}
        \end{algorithm}

        Algorithms \ref{Alg:Alg_fpg_lasso}-\ref{Alg:Alg_fpg_SM} require the Lipschitz constant $L_f$ of ${\bf M}$. This constant is readily given by noting that
            $$
                L_f=||{\bf M}||_2=\max_{i}\lambda_i,\;i=1,\ldots,N^2,
            $$
            with $\lambda_i$ being the $i$'th entry of ${\bm \lambda}$. The value $\max_{i}\lambda_i$ is calculated as part of Algorithm \ref{Alg:Alg6} and is given by
            $$
                L_f=\max_{i,j}b_{ij},\;i,j=1,\ldots,N,
            $$
            where $b_{ij}$ is the $ij$th entry of ${\bf B}$ from line 2 in Algorithm \ref{Alg:Alg6}.

        \subsection{Efficient calculation of {\bf v}}
        The vector ${\bf v}$ in (\ref{Eq:grad_f}) is the input data to Algorithms \ref{Alg:Alg_fpg_lasso}-\ref{Alg:Alg_fpg_SM}. Its $i$th element is given by
        $$
            v_i = {\bf a}_i^H{\bf R}_y(0){\bf a}_i,\;i=0,\ldots,N^2-1,
        $$
        with ${\bf a}_i$ representing the $i$th column of ${\bf A}$. Since ${\bf v}$ is an $N^2$ long vector, calculating its entries strictly by applying ${\bf A}^H$ and ${\bf A}$ on ${\bf R}_y(0)$ and taking the resulting diagonal is impractical as $N$ increases. Instead, it is possible to calculate its entries in two steps, as follows.

        The application of ${\bf a}_i$ on a matrix is very similar to the application of ${\bf A}$, only for a specific index $i$. We may write ${\bf a}_i$ more explicitly as
        $$
            {\bf a}_i={\bf H}({\bf f}_{k_i}\otimes {\bf f}_{l_i}),
        $$
        with $k_i=\floor*{{i\over N}}$ and $l_i=i \bmod N$. 
        By using the previously defined operations, ${\bf v}$ can be calculated as summarized in Algorithm \ref{Alg:Alg7}. This calculation needs to be performed only once, at the beginning of Algorithms \ref{Alg:Alg_fpg_lasso}-\ref{Alg:Alg_fpg_SM}. 
        
                \begin{figure}[ht!]
          \begin{subfigure}[b]{0.5\linewidth}
            \centering
            \includegraphics[trim={0cm 0cm 0cm 0cm},clip,width=0.8\linewidth]{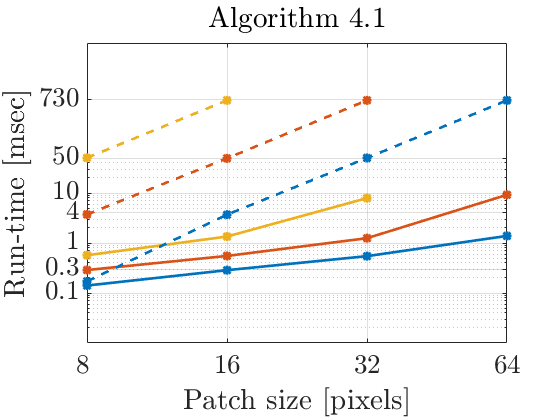}
            \caption{$l_1$ based recovery.}\label{Fig:Left}
            \vspace{4ex}
          \end{subfigure}
          \begin{subfigure}[b]{0.5\linewidth}
            \centering
            \includegraphics[trim={0cm 0cm 0cm 0cm},clip,width=0.8\linewidth]{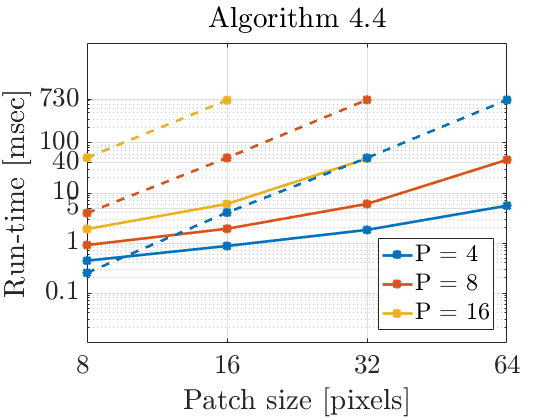}
            \caption{Wavelet based recovery.}\label{Fig:Right}
            \vspace{4ex}
          \end{subfigure}\vspace{-0.5cm}
          \caption{{\bf Algorithm run-time as a function of patch size.} Left panel shows run-time for a single iteration of Algorithm \ref{Alg:Alg_fpg_lasso}. Right panel shows run-time for a single iteration of Algorithm \ref{Alg:Alg_fpg_SM} (Daubechies wavelet filter with 8 taps and decomposition level of 2). Solid lines correspond to the frequency domain formulation (exploiting the BCCB structure of ${\bf A}^T{\bf A}$) while the dashed curves correspond to the spatial domain formulation. All values were averaged over $2000$ iterations. Vertical axis is in logarithmic scale, but the values are given in linear scale.}\label{Fig:RunTime}
        \end{figure}
        
        \begin{algorithm}
            \caption{Calculation of ${\bf v}$}
            \label{Alg:Alg7}
            \begin{algorithmic}
                \REQUIRE ${\bf H}$ and ${\bf R}_y(0)$
                \STATE {\bf Calculation of ${\bf Z}^H={\bf A}^H{\bf R}_y(0)$:}
                \STATE \hspace{0.3cm} {\bf 1:} Calculate ${\bf Q}={\bf H}^H{\bf R}_y(0)$
                \STATE \hspace{0.3cm} For each column of ${\bf Q}$, ${\bf q}_i$, $i=0,\ldots,M^2-1$:
                \STATE \hspace{0.6cm} {\bf 2:} Calculate ${\bf T}_i=N$\matst{ifft}$\{\Mscr({\bf q}_i)\}$ of length $N$
                \STATE \hspace{0.6cm} {\bf 3:} Calculate ${\bf E}_i=$\matst{fft}$\{{\bf T}^H_i\}$ of length $N$
                \STATE \hspace{0.6cm} {\bf 4:} Take the $i$th column of ${\bf Z}^H$ as $\Vscr({\bf E}^H_i)$
                \STATE {\bf Calculation of each element in ${\bf v}$:}
                \STATE \hspace{0.3cm} For each $i=0,\ldots,N^2-1$:
                    \STATE \hspace{0.6cm} {\bf 5:} ${\bf B}={\bf F}^H \Mscr({\bf H}^H{\bf z}_i)$, with ${\bf z}_i$ the $i$th column of ${\bf Z}$
                    \STATE \hspace{0.6cm} {\bf 6:} Calculate ${\bf u}={\bf F} {\bf b}_{l_i}$, with ${\bf b}_{l_i}$ the $l_i$th row of ${\bf B}$
                    \STATE \hspace{0.6cm} {\bf 7:} Take $v_i=u_{k_i}$, the $k_i$ entry of ${\bf u}$.
                \RETURN ${\bf v}$.
            \end{algorithmic}
        \end{algorithm}

\subsection{Algorithm run-time}
In this section, we compare the average run-time of our Fourier based formulation, i.e. using the structure depicted in Fig. \ref{Fig:MtrComp}b against the spatial domain formulation (Fig. \ref{Fig:MtrComp}a). Figure \ref{Fig:RunTime} shows the average run-time for a single iteration of Algorithm \ref{Alg:Alg_fpg_lasso} (left panel) and Algorithm \ref{Alg:Alg_fpg_SM} (right panel), performed on a $64$GB RAM, Intel i7-5960X@3GHz machine and implemented in Matlab (The Mathworks, Inc.). Each value is the average over $2000$ runs. The Eigenvalues of ${\bf M}$ and ${\bf v}$ are calculated a-priori. 

As expected, run-time increases as patch size increases, and as the value of $P$ increases. 
All curves are roughly linear, indicating an exponential growth in complexity as patch size increases, as the vertical axis is displayed in logarithmic scale (numerical values are given in linear scale). 
For the frequency domain formulation (solid lines), Fig. \ref{Fig:RunTime} shows that the execution time of each iteration is very fast for patches of sizes $8\times 8$, $16\times 16$ and $32\times 32$. On the other hand, run-time curves for the spatial domain formulation (dashed curves) are between one to two orders of magnitude higher, especially for higher values of $P$, such as $8$ and $16$. The value of $P$ needs to be increased as smaller features are needed to be resolved. These curves clearly motivate the use of our frequency domain formulation.
Moreover, it is recommended to divide the entire field of view to patches of $8-32$ pixels and process each patch independently. Since each patch is processed independently, the entire computational process can be parallelized for additional gain in efficiency.


\section{Simulations}
\label{Sec:Sim}

In this section, we provide further examples and characterization of SPARCOM. We start by providing an additional simulation to the results given in \cite{Solomon2018sparsity}, showing the ability of SPARCOM in recovering fine features absent in the diffraction limited movie, as well as providing additional comparisons to an improved SOFI formulation, termed balanced SOFI (bSOFI) \cite{geissbuehler2012mapping} and high emitter density STORM, implemented with the freely available ThunderSTORM software \cite{ovesny2014thunderstorm}. 
This sub-diffraction object and its corresponding SPARCOM recovery serves as a basis for an additional sensitivity analysis of SPARCOM to inexact knowledge of the PSF, presented in Fig. \ref{Fig:Fig6} and Fig. \ref{Fig:Fig7}. The next simulation presents the key advantages of SPARCOM in scenarios where assuming sparsity in other domains than the image domain leads to improved recovery results. We finish by providing experimental reconstruction results of SPARCOM, with our general super-resolution framework. 
These aspects complement the demonstration and analysis performed in \cite{Solomon2018sparsity}, thus providing a more comprehensive understanding of SPARCOM and its applications.

\subsection{Comparison of different super-resolution methods}
\begin{figure}[th!]
     \hspace{-1.1cm}
       \includegraphics[trim= 0cm 2cm 0cm 0cm, scale=0.45]{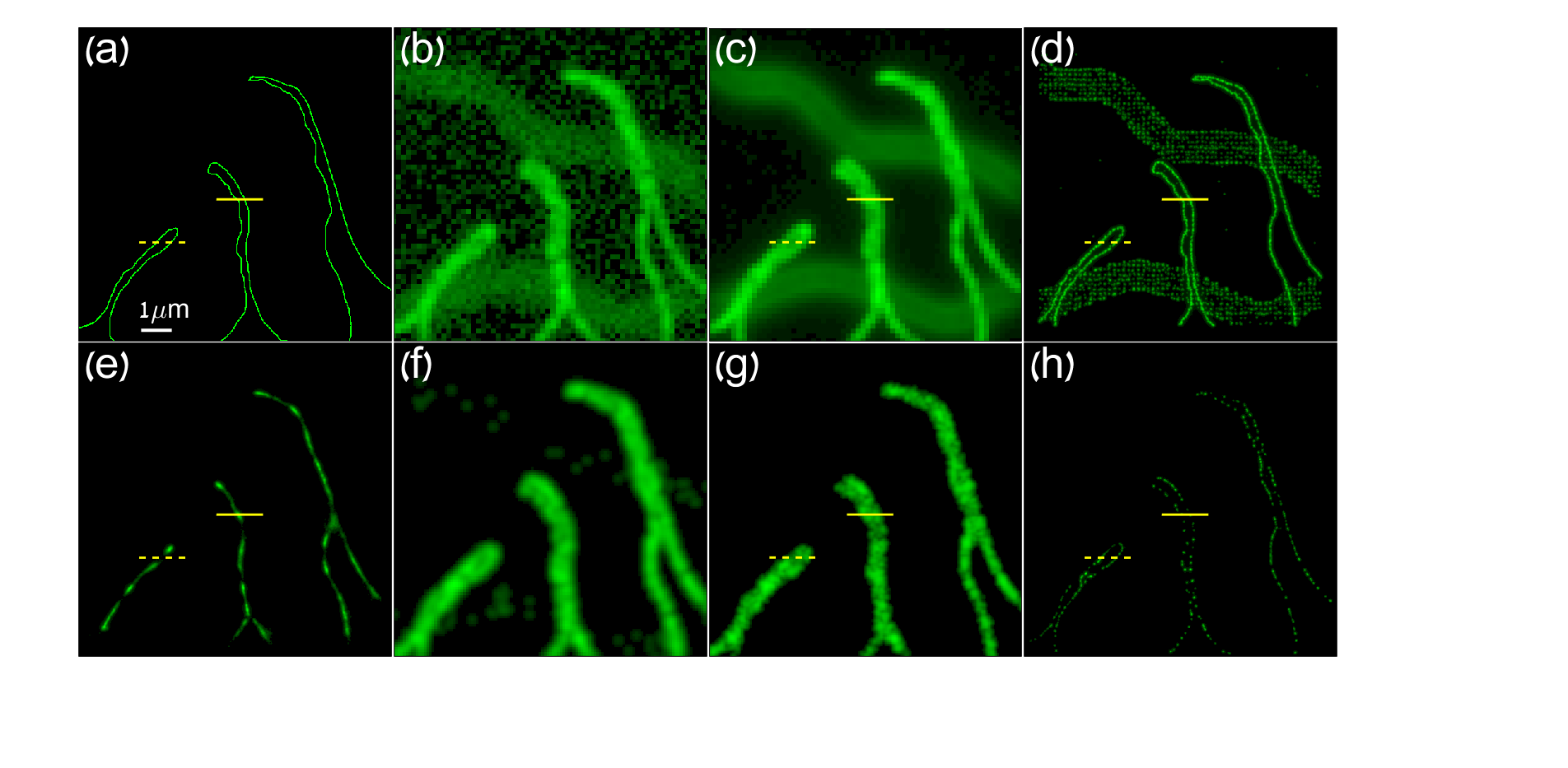}
       \caption{{\bf Reconstruction performance comparison of different methods.} {\bf Upper row:} (a) Ground truth: high resolution image of simulated sub-wavelength features. (b) Single diffraction limited frame from the movie, created by convolving the movie of fluctuating point emitters according to the locations in (a) with the PSF and adding Gaussian noise. (c)  Diffraction-limited image, taken by averaging all the frames in the movie. (d) ThunderSTORM recovery from 5000 low density frames. {\bf Lower row: recovered images from a noisy sequence of 1000 frames.} (e) Smoothed ThunderSTORM. (f) Correlations SOFI (zero time-lag). (g) $4^{th}$ order SOFI (in absolute value, zero time-lag). (h) SPARCOM recovery.}\label{Fig:Fig1}
     \end{figure}

     \begin{figure}[ht!]
     \centering
       \includegraphics[trim= 0cm 16cm 0cm 0.7cm, scale=0.35]{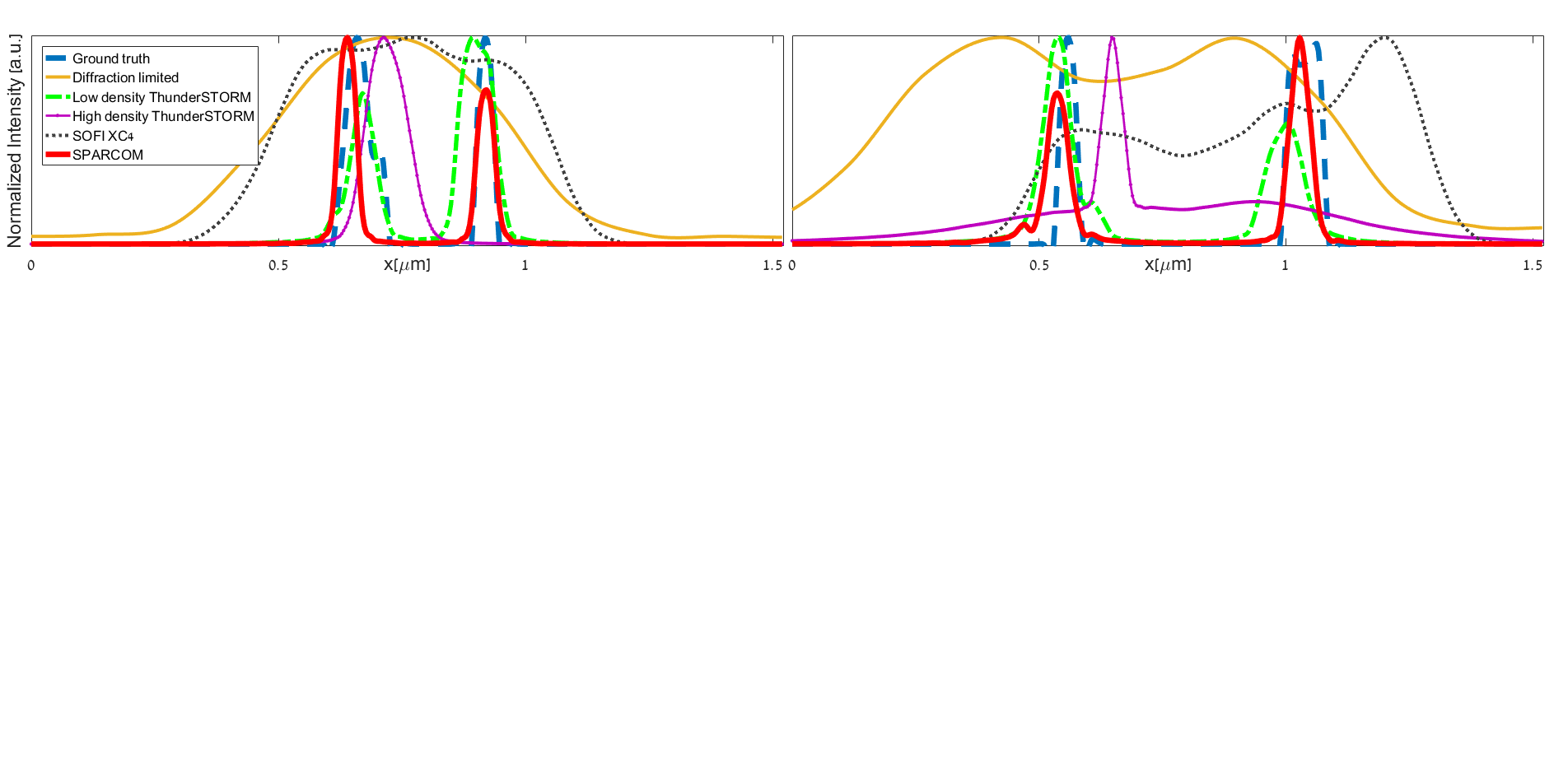} 
       \caption{Normalized cross-sections along the solid yellow line (left) and the dashed yellow line (right) of Fig. \ref{Fig:Fig1}, comparing the ground truth (dashed blue, Fig. \ref{Fig:Fig1}a), diffraction-limited image (solid yellow, Fig. \ref{Fig:Fig1}c), ThunderSTORM using 5000 low density frames (dash dot green, Fig. \ref{Fig:Fig1}d), ThunderSTORM using 1000 high density frames (solid thin purple, Fig. \ref{Fig:Fig1}e), $4^{th}$ order SOFI (dot black, Fig. \ref{Fig:Fig1}g), and SPARCOM (solid red, Fig. \ref{Fig:Fig1}h).}\label{Fig:Fig2}
     \end{figure}

    We numerically simulated a movie of sub-wavelength features 
     over 1000 frames, contaminated by additive Gaussian noise with $\textrm{SNR}=14.95$dB, 
     $$
        \textrm{SNR}=20\cdot\log_{10}{||{\bf Y}_{\textrm{movie}}||_F\over ||{\bf N}_{\textrm{movie}}||_F},
     $$
     were ${\bf Y}_{\textrm{movie}}$ is an $M^2\times T$ matrix, representing the entire blurred movie (each movie frame is column stacked as a single column in ${\bf Y}_{\textrm{movie}}$) and ${\bf N}_{\textrm{movie}}$ is the added noise to all the frames (same dimensions as ${\bf Y}_{\textrm{movie}}$). The movie also includes the simulation of out-of-focus filaments, which simulate unwanted fluorescence from objects outside the focal plane. Thus, they appear much wider than the in-focus simulated filaments. For both the in-focus and out-of-focus objects, we used the same Gaussian PSF, generated using the freely available PSF generator \cite{kirshner20133,kirshner20113d}, but with focal depths of $0nm$ and $1\mu m$, respectively. 

     In Fig. \ref{Fig:Fig1}a we show the simulated ground truth of the image with sub-wavelength features of size $512\times 512$ pixels. The imaging wavelength is $800nm$ with a numerical aperture of 1.4. We simulated two movies. The first is composed of 1000 high emitter density frames, while the second is composed of 5000 low emitter density frames of the same features. 

     Figure \ref{Fig:Fig1}b illustrates a single frame from the high density movie (each frame size is $64\times 64$ pixels and the pixel size corresponds to $160nm$), while Fig. \ref{Fig:Fig1}c shows the diffraction limited image (a sum of all $1000$ frames). The PSF (shown in Fig. \ref{Fig:Fig6}d after binning by a factor of $8$). 

     Figure \ref{Fig:Fig1}d shows a smoothed ThunderSTORM \cite{ovesny2014thunderstorm} reconstruction (freely available code) from the low emitter density movie. This image serves as a reference for the best possible reconstruction, when there are no temporal considerations. On the other hand, Fig. \ref{Fig:Fig1}e depicts smoothed ThunderSTORM reconstruction, performed with the high density movie of 1000 frames. Since the ground truth is of size $512\times 512$ pixels, the raw localizations image was resized to a $512\times 512$ image and smoothed with a Gaussian kernel. Figures \ref{Fig:Fig1}f and \ref{Fig:Fig1}g show the second and forth order SOFI images respectively (absolute values, zero time-lag). SOFI reconstructions were performed using the freely available code of bSOFI \cite{geissbuehler2012mapping}, which also includes a Richardson-Lucy deconvolution step with the discretized PSF used in our method. Last, Fig. \ref{Fig:Fig1}h displays the SPARCOM reconstruction ($512\times 512$ pixels) after smoothing with the same kernel used in Figs. \ref{Fig:Fig1}d and \ref{Fig:Fig1}e. Reconstruction was performed over $2000$ iterations and with $\lambda=10^{-3}$.

     Note that the SOFI reconstructions do not compare in resolution to the ThunderSTORM and SPARCOM recoveries. This additional comparison shows that, even when considering more advanced implementations of SOFI, such as bSOFI, the resolution increase does not match that of SPARCOM. Furthermore, it is evident that the SPARCOM recovery (Fig. \ref{Fig:Fig1}h) detects the ``cavities" within the hollowed features, similarly to low density ThunderSTORM (\ref{Fig:Fig1}d). When high emitters density is used, Fig. \ref{Fig:Fig1}e illustrates that ThunderSTORM recovery fails and no clear depiction of these features is possible.   

     In order to further quantify the performance of SPARCOM, Fig. \ref{Fig:Fig2} presents selected intensity cross-sections along two lines. In both profiles (solid and dashed yellow lines in the panels of Fig. \ref{Fig:Fig2}), several observations can be made. First, there is a good match between the locations and width of the SPARCOM (solid red) and low density ThunderSTORM (dash dot green) recoveries with the ground truth (dashed blue), indicating that SPARCOM achieves a comparable spatial resolution to ThunderSTORM, when there are no temporal constraints. Second, if temporal resolution is critical, i.e. it is essential to capture only a small number of high emitter density frames, then ThunderSTORM fails (solid thin purple), detecting only a single, misplaced peak, compared with the two peaks of the ground truth. Finally, in this scenario, SOFI reconstruction (dot black) failed in achieving good recovery.

     Figures \ref{Fig:Fig1} and \ref{Fig:Fig2} demonstrate that sparse recovery in the correlation domain achieves increased resolution with increased temporal resolution (5 times in this example) and detects the cavities within the sub-wavelength features which are absent in the low resolution movie, high density ThunderSTORM and SOFI reconstructions. This simulation adds upon the simulations presented in \cite{Solomon2018sparsity}, by comparing SPARCOM with bSOFI, which provides additional steps to the original SOFI scheme, such as a deconvolution step, as well as demonstrating the disadvantages of localization-based methods in the high density scenario (which can lead to a reduction in the total acquisition time).

\subsection{Super-resolution under different regularizers}
\begin{figure}[th!]
     \centering
     \hspace*{-1.1cm}
       \includegraphics[trim= 0cm 2cm 0cm 0cm, scale=0.45]{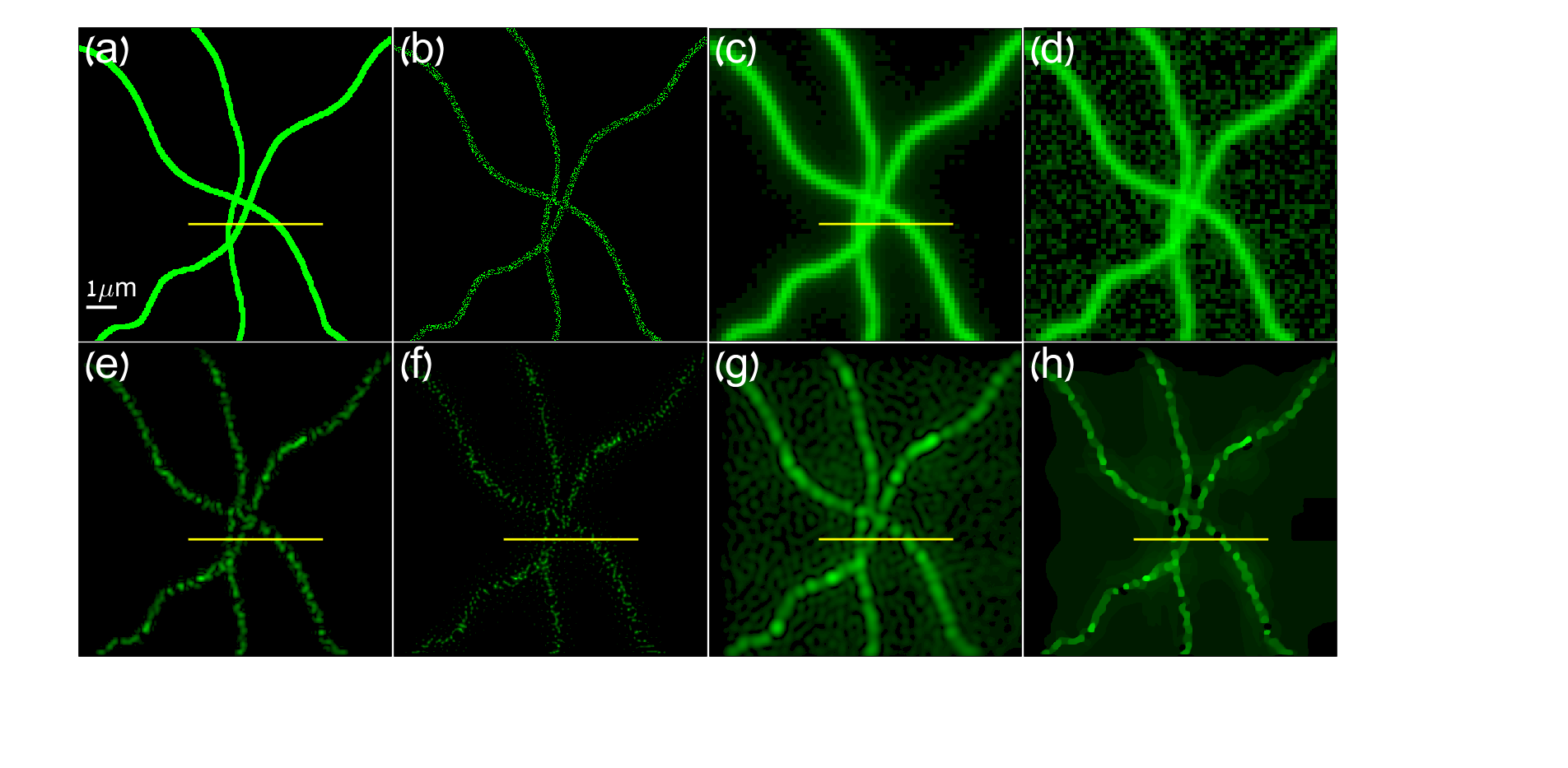}
       \caption{{\bf Regularized super-resolution.} {\bf Upper row: unprocessed data.} (a) Ground truth: high resolution simulated image. (b) Positions of emitters in the first frame. (c) Diffraction-limited image. (d)  Single diffraction limited frame.  {\bf Lower row: recovered images from a noisy sequence of 1000 frames.} (e) 2D wavelet reconstruction.  (f) $l_1$ reconstruction. (g) 2D DCT reconstruction. (h) Isotropic TV reconstruction.}\label{Fig:Fig4}
     \end{figure}

    Next, we tested our more general framework for super-resolution reconstruction. We simulated a movie of thick sub-diffraction filaments over 1000 frames with Gaussian noise ($\textrm{SNR}=17.72$dB). In Fig. \ref{Fig:Fig4}a we show the simulated ground truth of size $512\times 512$ pixels. The imaging wavelength is $800nm$ with a numerical aperture of 1.4. Figure \ref{Fig:Fig4}b shows the positions of the emitters for the first frame in the movie, while Fig. \ref{Fig:Fig4}c shows the diffraction limited image (a sum of all $1000$ frames). Figure \ref{Fig:Fig4}d shows a single frame from the simulated movie, where each frame size is $64\times 64$ pixels and the pixel size corresponds to $160nm$. We used the same PSF as before.

     \begin{figure}[ht!]
     \centering\hspace*{-1.5cm}
       \includegraphics[trim= 0cm 16cm 26.5cm 0.7cm, scale=0.6]{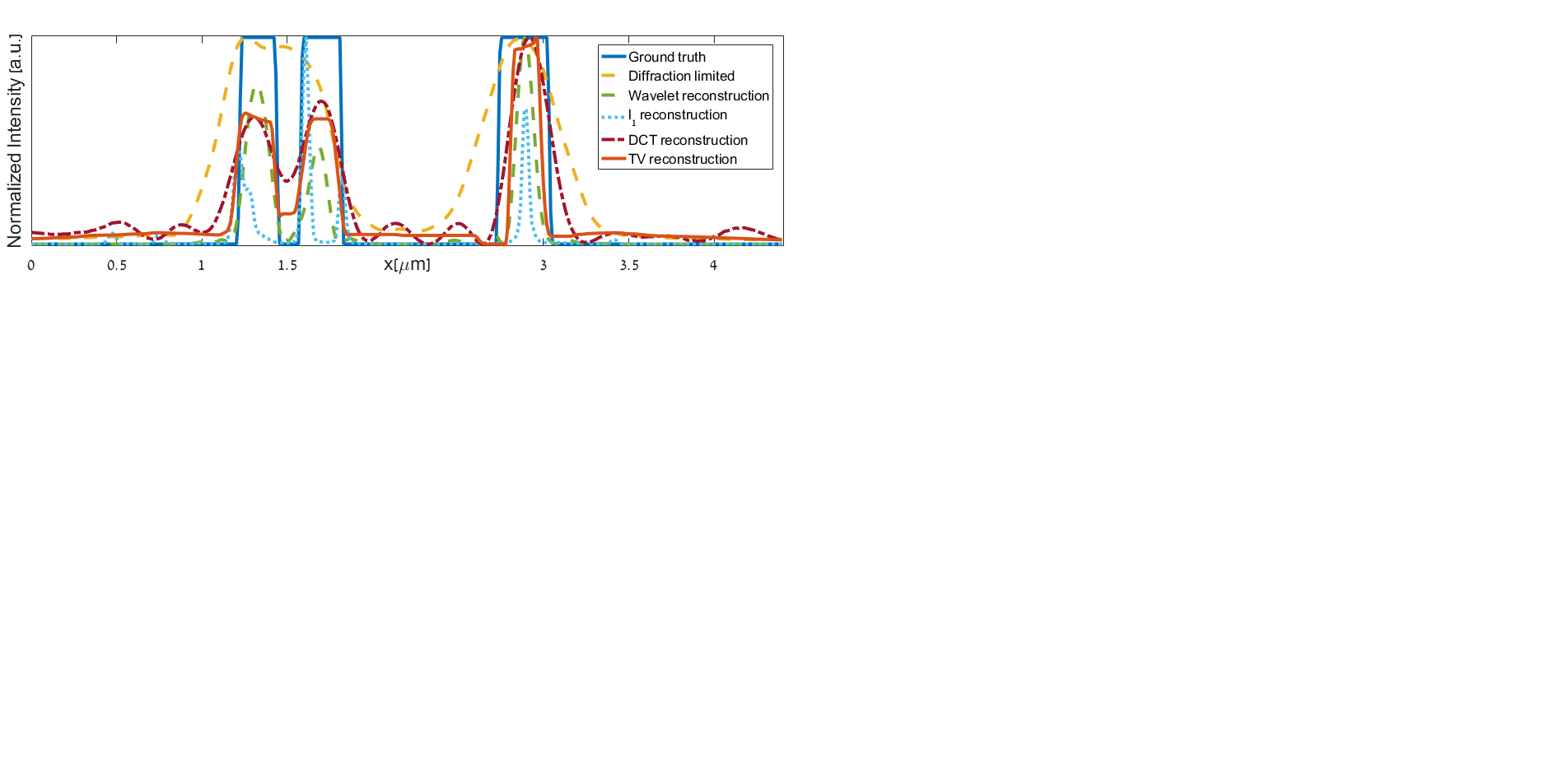} 
       \caption{Normalized cross-sections along the solid yellow line of Fig. \ref{Fig:Fig4}, comparing the ground truth (solid black, Fig. \ref{Fig:Fig4}a), diffraction-limited image (dashed yellow, Fig. \ref{Fig:Fig1}c), 2D wavelet reconstruction (dashed green, Fig. \ref{Fig:Fig1}e), $l_1$ reconstruction (blue dot, Fig. \ref{Fig:Fig1}f), 2D DCT reconstruction (dash-dot pink, Fig. \ref{Fig:Fig1}g), and isotropic TV reconstruction (orange solid x, Fig. \ref{Fig:Fig1}h).}\label{Fig:Fig5}
     \end{figure}

    Figure \ref{Fig:Fig4}e shows reconstruction in the 2D wavelet domain, while Fig. \ref{Fig:Fig4}f considers reconstruction under the assumption of a sparse distribution of molecules (Algorithm \ref{Alg:Alg_fpg_lasso}, $2000$ iterations, $\lambda=10^{-4}$ and smoothed with the same kernel as before). For the wavelet reconstruction we used Algorithm \ref{Alg:Alg_fpg_SM} with $2000$ iterations, $\lambda=8\cdot10^{-4}$ and $\mu=10^{-5}$. The wavelet and inverse-wavelet transform were implemented using the {\it Rice Wavelet Toolbox\footnote{http://dsp.rice.edu/software/rice-wavelet-toolbox}} V.3, with $2$ decomposition levels and a Daubechies scaling filter of $32$ taps produced by the function \matst{daubcqf} \cite{daubechies1988orthonormal}. Figure \ref{Fig:Fig4}g considers reconstruction in the 2D DCT domain, while Fig. \ref{Fig:Fig4}h shows reconstruction under an isotropic TV assumption. The DCT reconstruction used Algorithm \ref{Alg:Alg_fpg_SM} with $2000$ iterations, $\lambda=5\cdot10^{-4}$ and $\mu=10^{-5}$ and the isotropic TV recovery was performed using Algorithm \ref{Alg:Alg_fpg_TV} with $500$ iterations and $\lambda=10^{-4}$. Each denoising step (GP algorithm from \cite{beck2009fast}) used $100$ iterations.

    In Fig. \ref{Fig:Fig5} we show the normalized intensity profiles of the yellow lines in Fig. \ref{Fig:Fig4}, comparing the reconstruction performance of the various algorithms used previously. It is clear that the diffraction limited profile (dashed orange) conceals two filaments (solid black curve), which are distinguishable in all methods. However, the $l_1$ based reconstruction (i.e. sparsity assumption in the positions of the emitters) results in artifacts which gives the reconstructed image a grainy appearance and does not capture the true width of the filaments. On the other hand, the wavelet and TV based images show the filaments width more precisely, while DCT recovers a blurrier image of them. 

    Though this example is artificial, it serves to demonstrate that in some cases assuming sparsity in other domains than the original sparsity assumption may help produce reconstructions which are more faithful to the desired object and have smoother textures.

\subsection{Sensitivity of reconstruction to the PSF}
\begin{figure}[ht!]
     \centering\hspace*{-1.1cm}
       \includegraphics[trim= 0cm 2cm 0cm 0cm, scale=0.45]{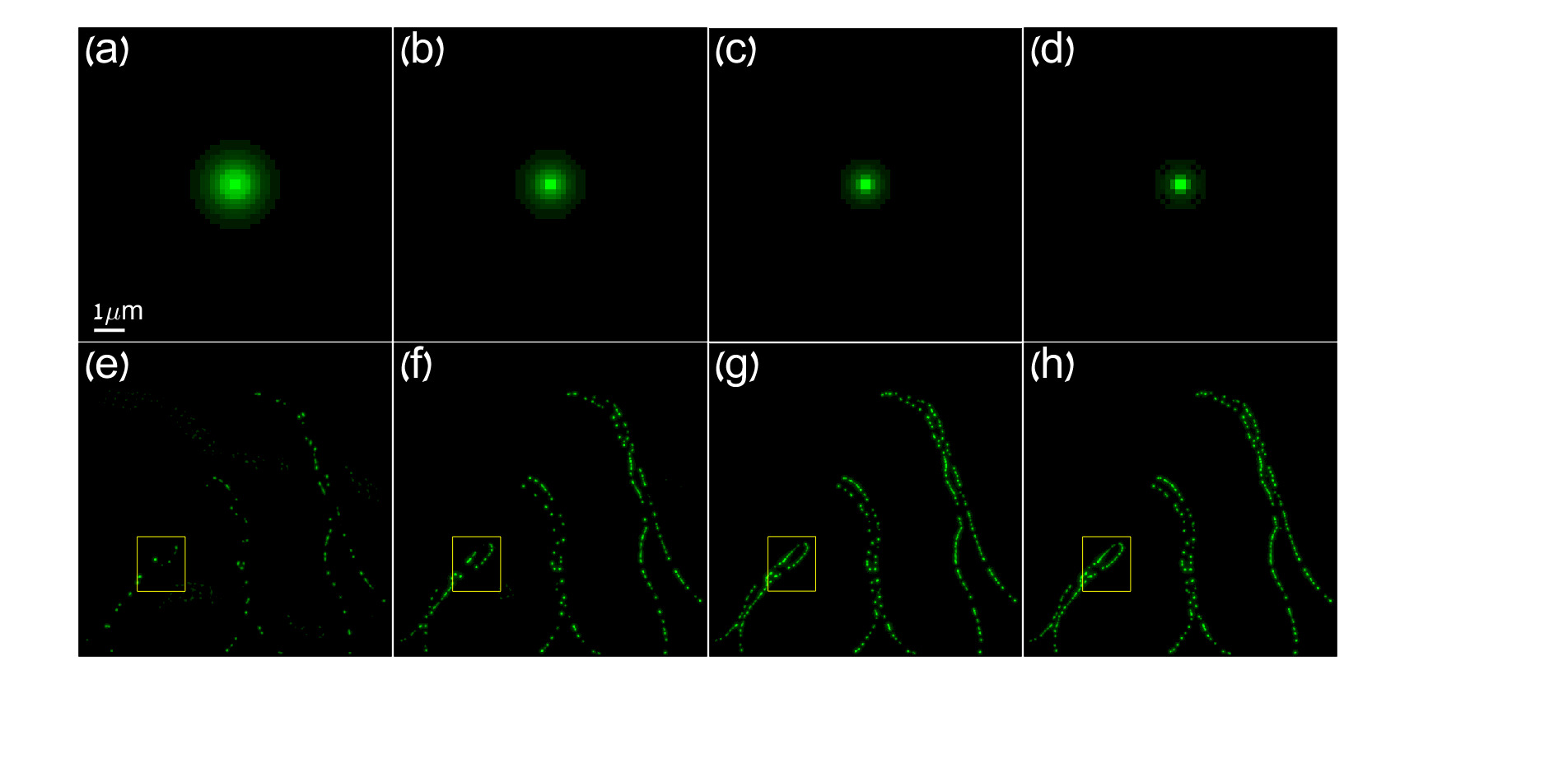}
       \caption{{\bf Reconstruction with out-of-focus PSFs.} {\bf Upper row:} (a)-(d) show simulated PSFs at distances $z=750,500,250,0$nm from the focal plane, respectively. The captured movie was generated with the PSF in (d). {\bf Lower row:} (e)-(h) show the reconstructed images using Algorithm \ref{Alg:Alg_fpg_lasso} for each of the PSFs (a)-(d), respectively. Yellow rectangles represent a magnified area shown in Fig.\ref{Fig:Fig7}.}\label{Fig:Fig6}
    \end{figure}

    Knowledge of the PSF is crucial for all the algorithms presented in this work. In practice, this knowledge is often imperfect and the PSF is usually estimated from the data \cite{Dertinger2010} or from a specific experiment used to determine it \cite{rietdorf2005microscopy}. When measuring the PSF of the microscope in an experiment, the position of the emitters or beads may not be exactly in the focal-plane, but rather a few hundreds of nanometers above or below it. Hence, we tested the reconstruction performance of Algorithm \ref{Alg:Alg_fpg_lasso} when used with different out-of-focus PSFs, to assess its sensitivity to inexact knowledge of the PSF. We used the same simulated data (and same SNR) as in Fig. \ref{Fig:Fig1} (which was generated with the PSF in Fig. \ref{Fig:Fig6}d) and simulated several PSFs measured at different distances from the focal plane. All reconstructions were performed over $2000$ iterations and with $\lambda=10^{-3}$. Figures \ref{Fig:Fig6}a-\ref{Fig:Fig6}d illustrate the different (binned) PSFs with varying distances from the focal plane, $z=750,500,250,0$nm, respectively. Each PSF was generated using the PSF generator, 
    and for $z=500$nm, the PSF width is twice the width of the in-focus PSF ($z=0$nm).

    \begin{figure}[ht!]
     \centering\hspace*{-1.1cm}
       \includegraphics[trim={0cm 2cm 0cm 11cm}, clip, scale=0.45]{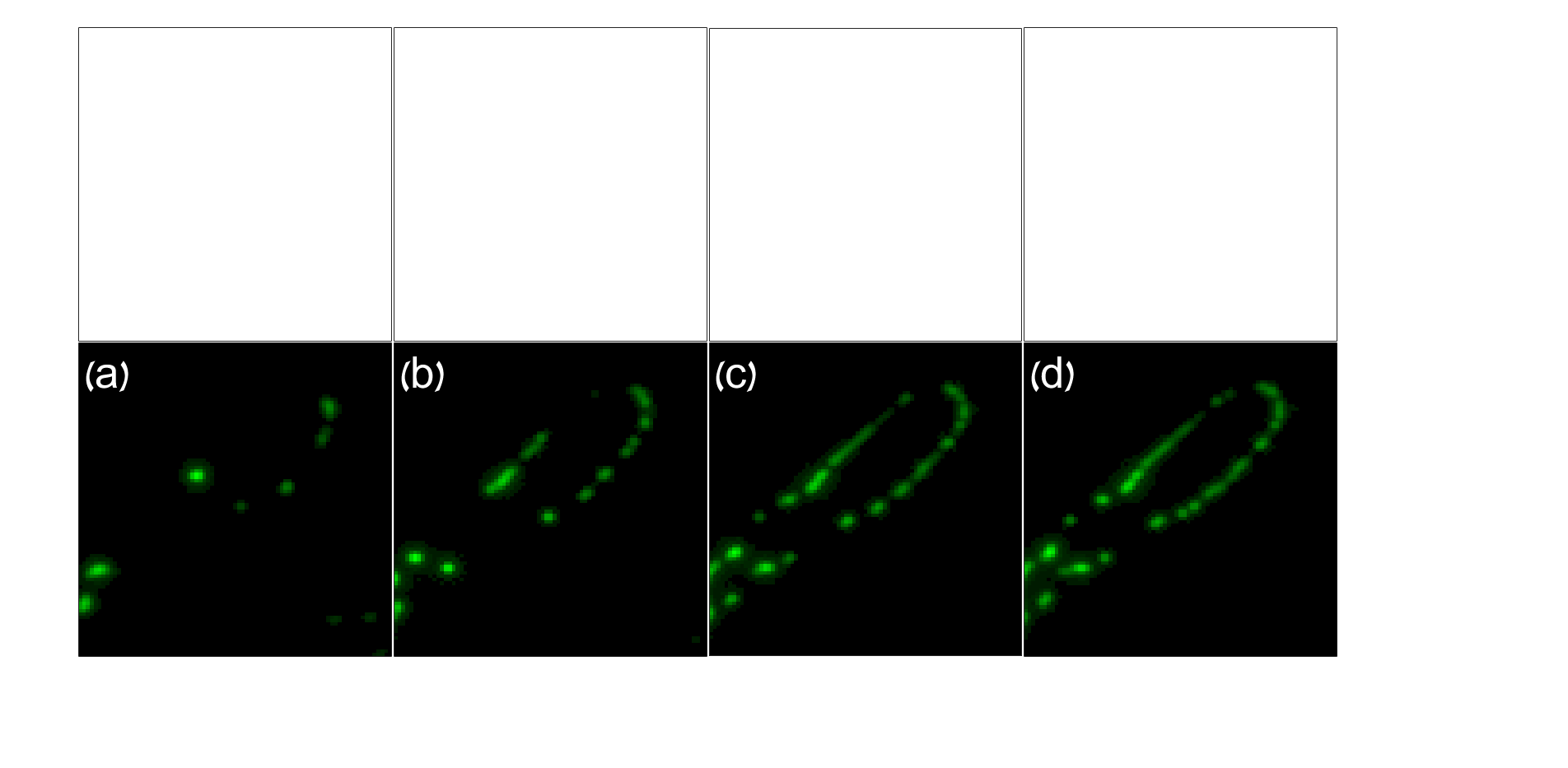}
       \caption{(a)-(d) show a zoom-in on the yellow rectangular windows in Fig. \ref{Fig:Fig6}(e)-Fig. \ref{Fig:Fig6}(h).}\label{Fig:Fig7}
    \end{figure}

    Figures \ref{Fig:Fig6}e-\ref{Fig:Fig6}h show the reconstruction results when used with the PSFs in Figs. \ref{Fig:Fig6}a-\ref{Fig:Fig6}d, respectively, while Figs. \ref{Fig:Fig7}a-\ref{Fig:Fig7}d show a zoom-in on the area inside the yellow rectangles in Figs. \ref{Fig:Fig6}e-\ref{Fig:Fig6}h. It is clear from both Figs. \ref{Fig:Fig6} and \ref{Fig:Fig7}, that as the PSF widens, reconstruction quality degrades, but similar reconstruction results in this example are given even for a PSF that twice as wide ($z=500$nm) as the in-focus PSF ($z=0$nm). This observation suggests that SPARCOM is fairly robust to inexact knowledge of the PSF, and deviations in its width (which correspond to deviations of several hundreds of nanometers in the axial depth of the PSF) can still lead to good reconstructions. 

\section{Experimental results}
In this section we further assess SPARCOM reconstructions on experimental dataset, under sparsity assumptions in different domains. In this example, applying Algorithm \ref{Alg:Alg_fpg_lasso} did not yield meaningful reconstruction, due to the width of the sub-diffraction features. Thus, we consider reconstruction under our generalized framework, and show that indeed performing sparse recovery with our generalized framework can clearly resolve sub-diffraction features from high-density movies. The dataset is freely available \cite{Min2014a} and consists of $160$ high-density frames of endoplasmic reticulum (ER) protein, fused to tdEos in a U2OS cell. The experimental setup consists of an imaging wavelength of $561$nm, numerical aperture of $1.3$ and pixel size of $100$nm. Each frame is $64\times 64$ pixels. The PSF was generated based on these acquisition parameters. We set $P=8$ and apply SPARCOM to this dataset, using the reconstruction algorithms presented in Section \ref{Sec:ProxGradSection}. 

\begin{figure}[ht!]
     \centering
       \includegraphics[trim= 0cm 0cm 0cm 0cm, scale=0.7]{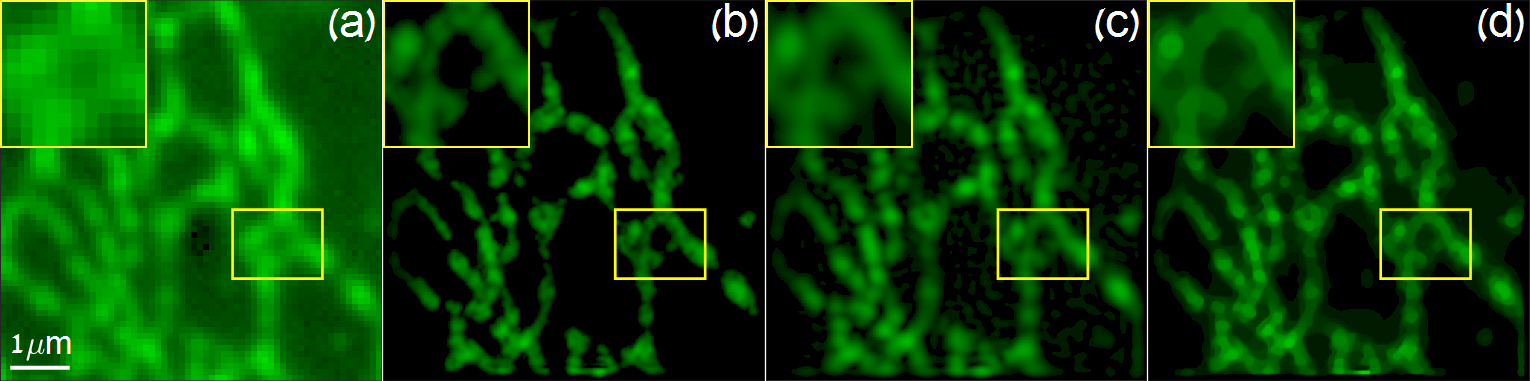}
       \caption{{\bf Experimental dataset results.} (a) Diffraction limited image composed of $160$ frames. (b) SPARCOM reconstruction in the wavelet domain. (c) SPARCOM recovery in the DCT domain. (d) SPARCOM reconstruction in the TV domain. Yellow insets indicate corresponding enlarged regions in the upper left corner of each panel.}\label{Fig:Exp}
    \end{figure}

Panel (a) of Fig. \ref{Fig:Exp} shows the diffraction limited image, while panels (b)-(d) show SPARCOM recoveries under the wavelet domain ($\lambda=2\cdot10^{-3}, \mu=10^{-5}$), DCT ($\lambda=, \mu=4\cdot10^{-4}, \mu=10^{-5}$) and isotropic TV ($\lambda=, \mu=10^{-4}$), respectively. All recoveries were performed with $2000$ iterations and wavelet recovery was performed using the {\it Rice Wavelet Toolbox} V.3 with $2$ decomposition levels and a Daubechies scaling filter of $32$ taps. Yellow insets indicate corresponding enlarged regions in the upper left corner of each panel. 

Considering panels (b)-(d), the wavelet-based reconstruction seems to be the sharpest, presenting sub-diffraction features which are absent in panel (a), while also depicting smooth and continuous features. DCT reconstruction seems to also present a smooth, albeit blurrier recovery. Finally, the TV-based image seems to also produce a consistent recovery, but with poorer resolution compared with the wavelet based recovery.

The enlarged regions show that all regularizers, but especially the wavelet regularizer can resolve sub-diffraction features which are completely absent in the diffraction limited image (left portion of the insets), while preserving a smooth depiction of the objects. This demonstration shows the benefit of performing recovery in additional domains, such as the wavelet domain, especially if the recovered features have an intricate morphology, of varying width.

\section{Discussion and Conclusions}
In this paper, SPARCOM, a method for super-resolution fluorescence microscopy with short integration time and comparable spatial resolution to state-of-the-art methods is further described and extended. By relying on sparse recovery and the uncorrelated emissions of fluorescent emitters, SPARCOM manages to reduce the total integration time by several orders of magnitudes compared to commonly practiced methods. We developed a thorough and detailed mathematical formulation of our method, and showed that considering reconstruction in the sampled Fourier domain results in a special structure of the gradient, which leads to a numerically efficient implementation relying of FFT operations. Moreover, we explored additional extensions of SPARCOM to scenarios in which assuming sparsity in other domains than simply the locations of the emitters leads to better recovery results. 

We conclude the paper by addressing the question of stable recovery of emitters in the noise-less and noisy cases, from a theoretical point of view. 
The authors of \cite{morgenshtern2016super} considered stable recovery of positive point sources from low-pass measurements. By solving a simple convex optimization problem in the noiseless case, they show that a sufficient condition for recovery is that $||{\bf x}||_0<M/2$, where $M$ is the number of low-pass measurements, without any regard to where the sources are on the high-resolution grid. That is, perfect recovery is possible although the measurement matrix ${\bf A}$ is highly coherent. In the presence of noise, such a condition is not sufficient and it is important to know how regular the positions of the emitters are, that is, how many spikes are clustered together within a resolution cell (see Definition 1 in \cite{morgenshtern2016super} for a proper definition of regularity). The bounds given in \cite{morgenshtern2016super} are with respect to specific theoretical PSFs. For example, the authors of \cite{bendory2016robust} found that the length of the resolution cell in the case of a 1D Gaussian kernel is $1.1/\sigma$, where $\sigma$ is the standard deviation of the Gaussian. 
    
    Since SPARCOM recovers the variance of each emitter, this scenario deals with the recovery of positive quantities, where now the desired signal is the variance of the emitters, and not their actual intensities. Thus, similar to the work of \cite{PP2015}, in the noiseless case we theoretically expect it to be possible to recover up to $O(M^2)$ emitter locations instead of $M/2$ for the same number of measurements. 

\section*{Acknowledgment}
We thank Prof. Shimon Weiss and Xiyu Yi for fruitful discussions on super-resolution fluorescence microscopy in general, and SOFI in particular.


%

\appendix
\section{Proof of matrix {\bf M} being BCCB}
\label{Seq:append}
We begin by defining {\it circulant} and {\it block circulant with circulant blocks} (BCCB) matrices  \cite{hansen2006deblurring}, \cite{gray2006toeplitz}.
\begin{defn}
  A matrix ${\bf C}\in\bbC^{N\times N}$ is said to be {\it circulant} if
  $$
    C_{ij}=c_{(j-i)\hspace{-0.2cm}\mod N},\;\forall i,j=1,\ldots,N,
  $$
  for some $c_{(\cdot)}\in \bbC$, where $C_{ij}$ is the $ij$th entry of ${\bf C}$.
\end{defn}
\begin{defn}
  A matrix is said to be {\it block circulant with circulant blocks} if it can be divided into $N\times N$ square blocks, where each block is circulant and the matrix is circulant with respect to its blocks, e.g.:
  \begin{equation}
  \label{Eq:BCCB_example}
    {\bf B} = \left[\begin{array}{cccc}
        {\bf C}_0        & {\bf C}_{N-1} & \hdots  & {\bf C}_1 \\
        {\bf C}_1        & {\bf C}_0        & \hdots  & {\bf C}_2 \\
        \vdots             & \vdots             & \ddots & \vdots\\
        {\bf C}_{N-1} & {\bf C}_{N-2}  & \hdots & {\bf C}_0\\
    \end{array}\right],
\end{equation}
where each ${\bf C}_i,\;i=0,\ldots,N-1$ is an $N\times N$ circulant matrix.
\end{defn}
A circulant matrix of size $N\times N$ is completely defined by its first column vector, and so has $N$ degrees of freedom. Similarly, a BCCB matrix of size $N^2\times N^2$ is completely defined by its first column and has $N^2$ degrees of freedom. Denote the first column of ${\bf B}$ as ${\bf b}\in\bbC^{N^2}$, such that its $i$th element is denoted by $b_i$. In the following proof, we will show that the general element of ${\bf B}$ (and ${\bf M}$) can be represented by two independent sets of indices, the first corresponding to block circularity between $N\times N$ blocks and the second corresponding to circularity of the entries within each block. These two sets of indices correspond to partitioning ${\bf b}$ into $N$ non-overlapping vectors, each of length $N$, the first set indicates which is the right partition and the second to the right element within that partition. For the general element of (\ref{Eq:BCCB_example}), this property can be written more explicitly as,
\begin{equation}
    \label{Eq:bccb_gen_el}
    B_{ij}=b_{\left((k_j-k_i)\hspace{-0.2cm}\mod N\right)\cdot N+(l_j-l_i)\hspace{-0.2cm}\mod N},
\end{equation}
with $b_{(\cdot)}\in\bbC$, $i=k_iN+l_i,\;i=0,\ldots,N^2-1$ (same for the index $j$), such that $l_i,l_j=0,\ldots,N-1$ correspond to the position of (\ref{Eq:bccb_gen_el}) inside an $N\times N$ circulant block, and $k_i,k_j=0,\ldots,N-1$ correspond to one of the $N\times N$ blocks of ${\bf B}$. Notice that by the above construction, the values of $k_i$ and $k_j$ are increased by one, every $N$ increments of $l_i$ and $l_j$. 

We now prove that ${\bf M}$ is a BCCB matrix. 
\begin{proof}
Recall that ${\bf M}=|{\bf A}^H{\bf A}|^2$ and that $|\cdot|^2$ is performed element-wise. 
We start by considering the structure of ${\bf A}^H{\bf A}$:
\begin{equation}
\label{Eq:AHA}
  {\bf A}^H{\bf A}=({\bf F}_M^H\otimes{\bf F}_M^H){\bf H}^H{\bf H}({\bf F}_M\otimes{\bf F}_M),
\end{equation}
with ${\bf F}_M$ being a partial $M\times N$ discrete Fourier matrix (its $M$ rows are the corresponding $M$ low frequency rows from a full $N\times N$ discrete Fourier matrix) and ${\bf H}$ an $M^2\times M^2$ diagonal matrix. Denoting the $m$th column of ${\bf F}_M^H\otimes{\bf F}_M^H$ by $\tilde{{\bf f}}_m,\;m=1,\ldots,M^2$, we may write (\ref{Eq:AHA}) equivalently as 
\begin{equation}
\label{Eq:AHA2}
  {\bf A}^H{\bf A}=\dsum_{m=1}^{M^2}h_{m}\tilde{{\bf f}}_m\tilde{{\bf f}}_m^H,
\end{equation}
with $h_{m}$ the $m$th entry diagonal element of ${\bf H}^H{\bf H}$. 

The $m$th column of ${\bf F}_M^H\otimes{\bf F}_M^H$ is the Kronecker product of two columns from ${\bf F}_M^H$, say $\hat{{\bf f}}_{m_1}$ and $\hat{{\bf f}}_{m_2},\;m_1,m_2\in 0,\ldots,M-1$, where 
$$
  \hat{{\bf f}}_m=[1,e^{j{2\pi\over N}m},\ldots,e^{j{2\pi\over N}m(N-1)}]^T,\;m=0,\ldots,M-1.
$$
Replacing the summation over $m$ with a double sum over $m_1$ and $m_2$, (\ref{Eq:AHA2}) can be written more explicitly as 
\begin{equation}\label{Eq:AHA_app}
  {\bf A}^H{\bf A}=\dsum_{m_1=0}^{M-1}\dsum_{m_2=0}^{M-1}(\hat{{\bf f}}_{m_1}\otimes \hat{{\bf f}}_{m_2})(\hat{{\bf f}}_{m_1}^H\otimes \hat{{\bf f}}_{m_2}^H)h_{(M\cdot m_1+m_2)}.
\end{equation}
%

The $ij$th element of ${\bf M}$ is derived directly from (\ref{Eq:AHA_app}) and has the form 


\small
\begin{equation}
  \label{Eq:BCCB_prf}
  \begin{array}{l}
  M_{ij}=\left| \dsum_{m_1=0}^{M-1}\dsum_{m_2=0}^{M-1}e^{j{2\pi\over N}m_1(k_j-k_i)}e^{j{2\pi\over N}m_2(l_j-l_i)} h_{(M\cdot m_1+m_2)}\right|^2,\;
  i,j=0,\ldots,N^2-1,
  \end{array}
\end{equation}
\normalsize
where $k_i=\floor*{{i\over N}}$ and $l_i=i \bmod N$ (also for the index $j$). Note that the value of $k_i$ changes only between each $N\times N$ block, while the value of $l_i$ changes between the entries of each block. This construction directly implies that $i=k_iN+l_i$ (also for $j$), as indicated in (\ref{Eq:bccb_gen_el}). 





It can now be observed that $M_{ij}$ is composed of two independent sets of indices. Since ${\bf M}$ is of size $N^2\times N^2$, we can divide it to $N\times N$ non-overlapping blocks of the same size (similar to Fig. \ref{Fig:MtrComp}, right panel). The first exponential term $e^{j{2\pi\over N}m_1(k_j-k_i)}$ corresponds to $M$ being a block circulant matrix with $N\times N$ blocks. This can be seen by the construction of $k_i$ and $k_j$, since $k_j-k_i\in[-(N-1),\ldots,N-1]$ and by the periodicity by $N$ of the exponential term. 

The second set of indices, $l_j-l_i$ corresponds to each $N\times N$ block being circulant. This can be seen by the term $e^{j{2\pi\over N}m_2(l_j-l_i)}$, since $l_j-l_i\in[-(N-1),\ldots,N-1]$ and due to the periodicity by $N$ of the exponent. 
Thus, $M_{ij}$ has a structure similar to (\ref{Eq:bccb_gen_el}), with two independent sets of indices, the first corresponds to block circularity and the second to circularity within each block.
Consequently, ${\bf M}$ is a BCCB matrix.
\end{proof}
\ifCLASSOPTIONcaptionsoff
  \newpage
\fi


%


\bibliographystyle{ieeetran}
\bibliography{Bib_Mendeley}

\end{document}